\documentclass[final,1p,times]{elsarticle}
\usepackage{graphicx}
\usepackage{booktabs}

\usepackage{epstopdf}
\usepackage{amsmath}
\usepackage{amssymb}
\usepackage[english]{babel}
\usepackage[svgnames]{xcolor}

\newcommand{\bld}[1]{\boldsymbol{#1}}

\journal{Nuclear Physics A}

\begin{document}

\graphicspath{{figs/}}

\begin{frontmatter}

  \title{Fission barriers and probabilities of spontaneous fission
    \\ for elements with Z$\geq$100}

  \author[ab]{A.~Baran}
  \author[mk]{M.~Kowal} \author[pgr]{P.-G.~Reinhard}
  \author[lr]{L.M.~Robledo} \author[as]{A.~Staszczak}
  \author[mw]{M.~Warda}

  \address[ab,as,mw]{University of Maria-Curie-Sk{\l}odowska, 20031 Lublin,
    Poland}

  \address[mk]{National Centre for Nuclear Research, Ho\.za 69,
    PL-00-681 Warsaw, Poland}

  \address[lr]{Departamento de F\'isica Te\'orica, Universidad
    Aut´onoma de Madrid, 28049 Madrid, Spain}

  \address[pgr]{Institute of Theoretical Physics, University of
    Erlangen-Nuremberg, Germany}

  \begin{abstract}
    This is a short review of methods and results of calculations of
    fission barriers and fission half-lives of even-even superheavy
    nuclei.  An approvable agreement of the following approaches is
    shown and discussed: The macroscopic-microscopic approach based on
    the stratagem of the shell correction to the liquid drop model and
    a vantage point of microscopic energy density functionals of
    Skyrme and Gogny type selfconsistently calculated within
    Hartree-Fock-Bogoliubov method.  Mass parameters are calculated in
    the Hartree-Fock-Bogoliubov cranking approximation.  A short part
    of the paper is devoted to the nuclear fission dynamics.  We also
    discuss the predictive power of Skyrme functionals applied to key
    properties of the fission path of $^{266}$Hs. It applies the
    standard techniques of error estimates in the framework of a
    $\chi^2$ analysis.
  \end{abstract}


\end{frontmatter}

\section{\label{sec:intro}Introduction}

Considerable progress in the experimental synthesis of heaviest nuclei
has been achieved recently by the Flerov Laboratory in Dubna
\cite{Ogan99a,Ogan99b,Ogan00a,Ogan00b,Ogan04,Ogan10,Ogan06,Ogan09,Ogan12,Liu13,Ogan13}
and was partially confirmed in the laboratories at GSI
\cite{Hes97,Mun84,Hof98,refId0,Dul10}, LBNL \cite{Stav09} and RIKEN
\cite{Haba11}. Nonetheless, the fundamental question of what is the
largest possible atomic number of an atomic nucleus is still
unanswered. For example, at the time of writing this manuscript the
heaviest identified super-heavy element is $Z=118$, but even larger
elements are not yet excluded. Due to substantial Coulomb repulsion
for super-heavy nuclei the stability decreases rapidly when going up
in system size. Only shell effects allow super-heavy (SH) nuclei with
$Z\ge100$ to survive.  There are two dominant processes of
disintegration for such heavy systems: alpha particle emission and
fission. It is the subject of this contribution to provide an overview
of the state of the art relating to the latter process, fission of SH
nuclei. Two different microscopic methods for estimating the stability
of SH nuclei are used. There are, on the one hand, the self-consistent
approaches using effective interactions as, e.g., the Hartree-Fock
(HF) plus BCS (HFBCS) or Hartree-Fock-Bogoliubov (HFB) method, and, on
the other hand the more phenomenological macroscopic-microscopic (MM)
method. The latter is less demanding and thus allows even extensive
studies of multidimensional fission landscapes.  Although these models
differ quantitatively, they agree in predicted topological properties
as, e.g., the prolate deformed SH nuclei with Z = $100-112$ and
N$\le170$, which are confirmed experimentally for nuclei around
$^{254}$No \cite{Reit99} and spherical or oblate deformed systems
with N$=174-184 $ (see \cite{Cwiok96npa,Cwiok:05}).

The height of the fission barrier, $B_{f}$, is one of the most
important ingredients to calculate the survival probability of SH
elements synthesised in heavy-ion reactions. It allows to estimate the
competition between the fission process and particle emission. To
evaluate $B_{f}$, one needs the energy as function of collective
deformation, the potential energy surfaces (PES). Ideally, one has to
figure out all competing minima and saddle points on multidimensional
maps.  Having preconceived ideas about the expected fission landscape,
one usually one restricts the search to a one-dimensional fission
path.  In any case, the PES is the basic ingredient for the evaluation
of fission half-life (HL).

The last review of theoretical properties of SH elements covering MM
and self-consistent models was published eight years ago~\cite{SP07}.
Since then substantial progress has been made in the development of
models and methods especially by Lublin-Madrid
collaboration~\cite{Warda11,Warda:2012aa}, Lublin-Oak Ridge-Warsaw
group~\cite{Staszczak-13,Sheikh:2009aa,Pei:2009aa,Baran11c,Jhilam13,Jhilam14},
and Erlangen-Darmstadt team~\cite{Sch09a} -- energy density functional
(EDF) HFB approaches -- and in the MM model by the
Warsaw~\cite{Kowal2010a,Kowal2009,Sobiczewski2006}, Los Alamos
school~\cite{Moller2009} and the relativistic mean-field (RMF)
\cite{Abu2012}.

In this paper, we are discussing methods to obtain the PES, barrier
heights, mass parameters and fission HLs using modern
approaches like the shell correction method in a multidimensional
space of deformations and/or the method of energy density functionals
(EDF) of Skyrme and Gogny types treated self-consistently in the HFB
framework.
The main purpose is a comparison of results of these
different approaches and the discussion of uncertainties that affect
the data in the process of calculation. 

The paper is outlined as follows: Section~\ref{sec:mm} deals with
MM methods and data. In Section~\ref{sec:scm}, we
deliver the discussion of microscopic Skyrme's and Gogny's functionals
approaches and results.  The short Section~\ref{sec:mass} tackles the
issue of mass parameters and their effect on the HLs. 
Section~\ref{sec:compar} relates the calculated data to the
experimentally known cases.  Section~\ref{sec:dyn} deals with an
influence of dynamical treatment of fission process based on the
action minimisation on fission half-lives.  Section~\ref{sec:err}
discusses an estimate of extrapolation errors for the various
observables.  Conclusions are given in Section~\ref{sec:summary}.

\section{Predictions for barriers and life-times}

\subsection{\label{sec:mm}Macroscopic-microscopic (MM) approach}
The MM model composes the total energy of the nucleus from two
complementary parts. One is the macroscopic energy, usually some
version of a liquid drop (LD) energy and the other the shell
correction which describes the influence of single particles (SP)
energies and their quantum shell structure beyond the LD.  Both parts
are modelled independently. The LD part is taken from the extensive
fits within the LD model.  The results discussed here were obtained
within a MM model based on the single particle (SP) spectrum of a
deformed Woods-Saxon potential \cite{WS}. The SP spectrum obtained
from the deformed Woods-Saxon potential is used as input to compute
the Strutinsky shell correction energy \cite{str67,str68}. For the
macroscopic part we used the Yukawa plus exponential model~\cite{KN}.
The pairing parameters and three parameters of the macroscopic energy
formula were determined in \cite{WSpar} by a fit to masses of
even-even nuclei with $Z\geq84$ and $N>126$ as given in
\cite{Wapstra}.

\begin{figure}[th]
  \centerline{\includegraphics[width=.8\textwidth]{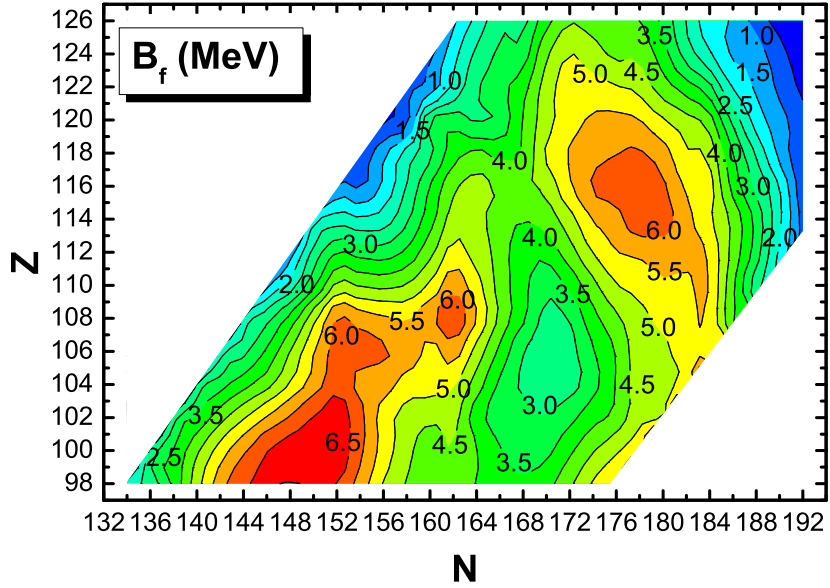}}
  \caption{{\protect Contour map of calculated fission barrier heights
      $B_{f}$ for even-even superheavy nuclei.
      \label{fig:Bftot}}}
\end{figure}

In Figure \ref{fig:Bftot} we display the height of the first fission
barriers $B_{f}$ for these nuclei calculated within the MM model as
the difference between the lowest saddle point energy and the ground
state one $E_{I}$. It can be seen that in the whole region of
considered nuclei the barriers are smaller than 7 MeV. The highest
values are obtained for the nuclei $^{270}108_{162}$,
$^{292}114_{178}$ and around the nucleus $Z\approx100, N\approx150$.
\begin{figure}[!h]
  \centerline{
    \includegraphics[width=0.48\linewidth]{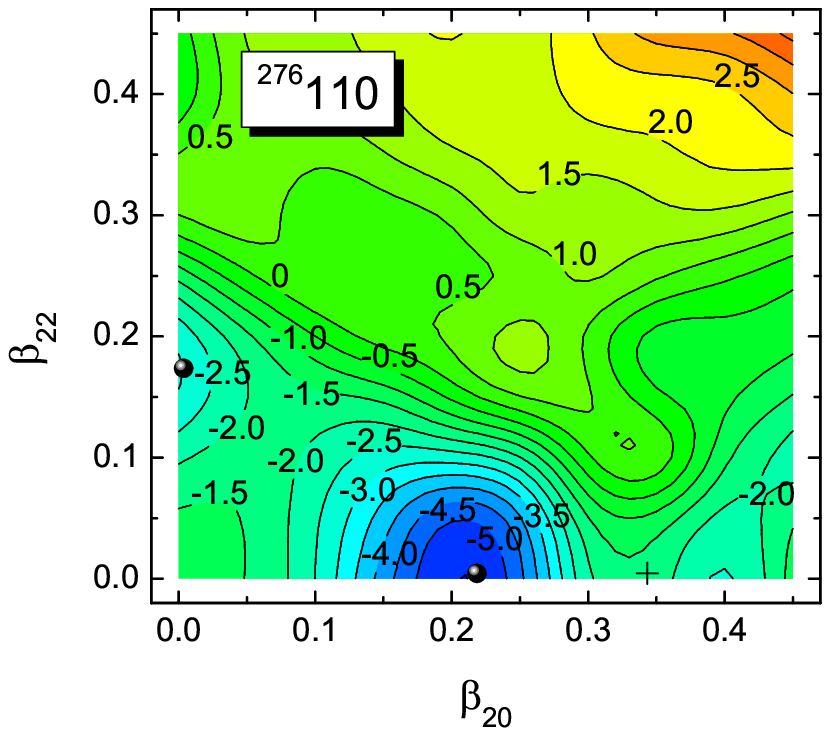}
    \includegraphics[width=0.48\linewidth]{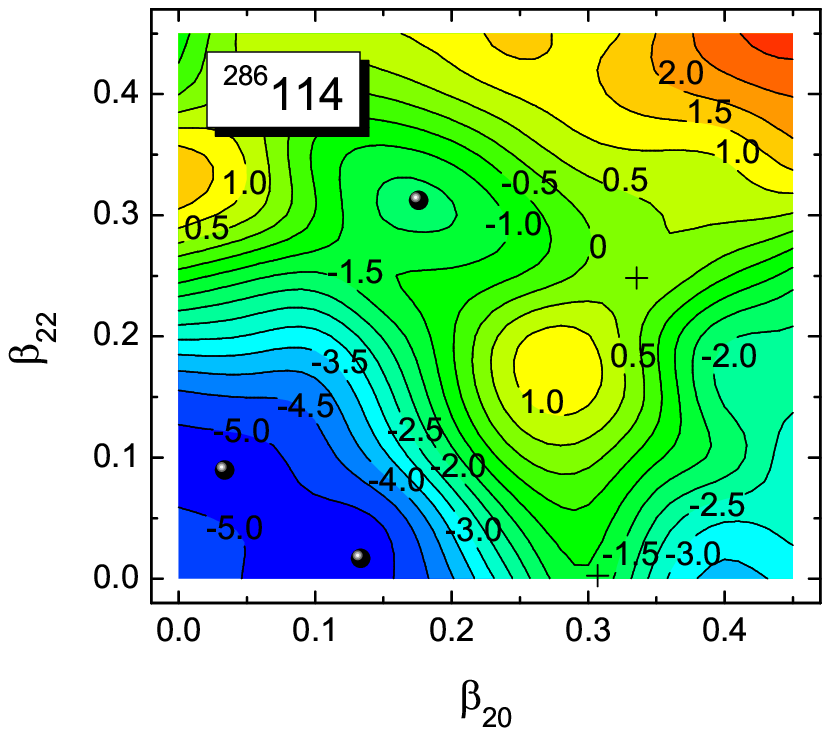}}
  \centerline{
    \includegraphics[width=0.48\linewidth]{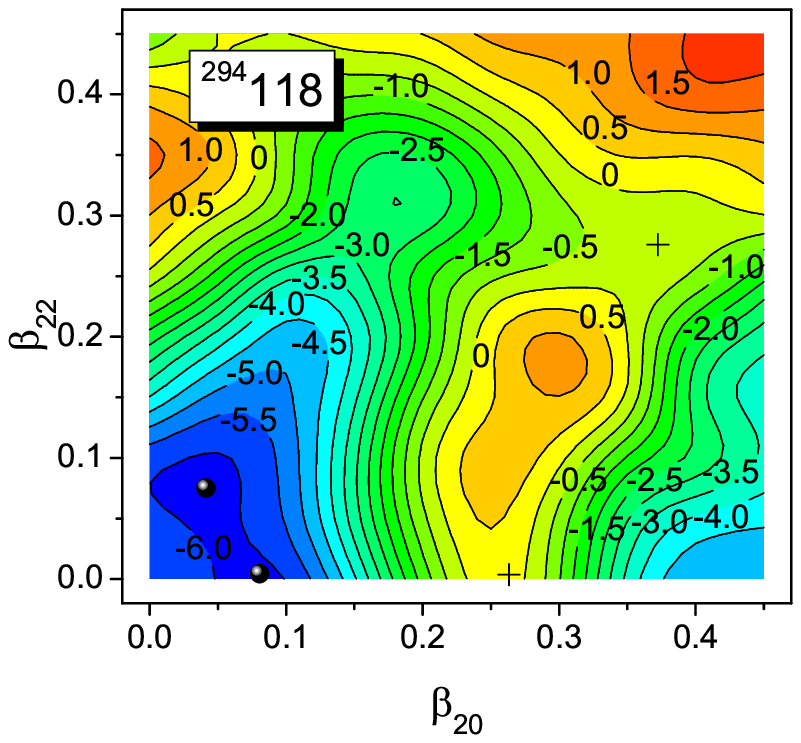}
    \includegraphics[width=0.48\linewidth]{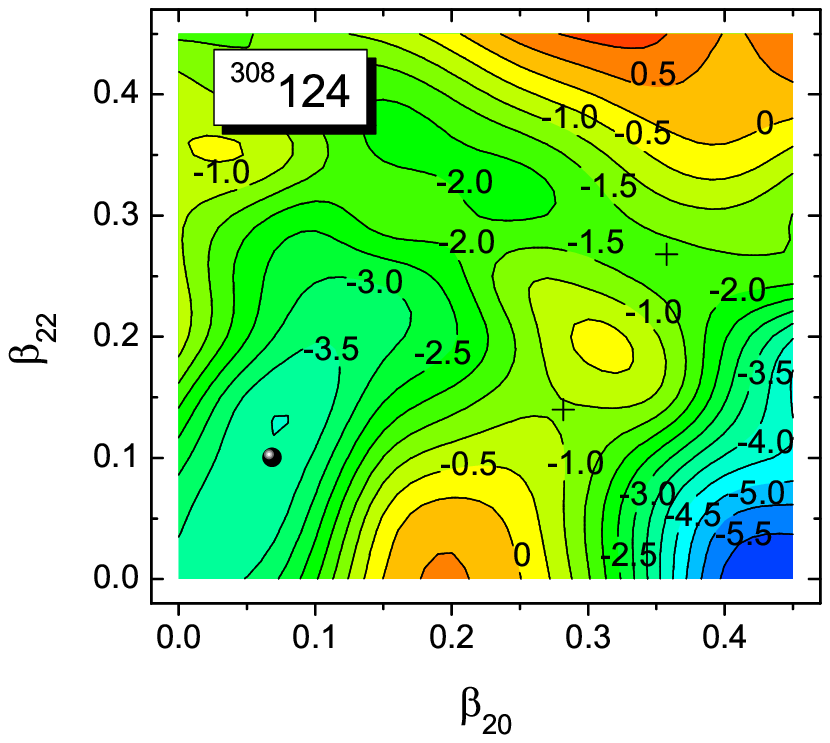}}
  \caption{Energy surfaces: $E-E_{mac}(sphere)$ for the three of
    experimentally detected evaporation residues and one hypothetical
    heavier system.}
  \label{fig:PESMM}
\end{figure}

Another important observation concerns the behaviour of fission barrier
heights with increasing proton number of these heaviest nuclei. One
can recognise that the quite high barrier for $^{296}118_{178}$
rapidly decreases reaching a value $1.43$ MeV for the
$^{312}126_{186}$ nucleus. Obviously such a compound system does not
have any chance to survive against fission in the MM model. This is in a
contradiction to self consistent models, according to fission barrier
for neighbouring nucleus is about $12$ MeV \cite{STASZ05, STASZ07}.

Fission barriers for even-even SH elements \cite{Kowal2010a} obtained
within the Woods-Saxon (WS) model have characteristic maxima at
$Z=108$, $N=162$ (deformed magic shells) and $Z=114$, $N=178$ (not
184). For the latter system the barrier reaches 6.34 MeV and then
decreases with $Z$.  This is in contrast with Skyrme SkM* predictions
\cite{Staszczak-13} of the highest barriers for $Z=120$
\cite{Staszczak-13}, which is related to the proton magic gap.  One
has to emphasise that the barriers from the WS model are in good
agreement with data for even-even actinides: for 18 first and 22
second barriers the rms deviation equals 0.5 and 0.7 MeV, respectively
\cite{Kowal2010a,Kowal2010b}.

Potential energy surfaces (PES) are often calculated and projected on
the triaxial ($\beta \cos\gamma$, $\beta \sin\gamma$) plane.  Examples
of energy maps in ($\beta \cos\gamma$, $\beta \sin\gamma$) plane,
necessary to evaluate fission barriers are shown in
Fig. \ref{fig:PESMM} (The total energy is normalised in such a way
that its macroscopic part is equal zero at the spherical shape of a
nucleus.)

Calculations of FRDM presented in~\cite{Moller15} confirm
aforementioned findings of the WS model e.g. region of the highest
fission barrier around $^{270}$Hs and $^{294}$Lv$_{178}$, decrease of
stability between these two maxima as well as beyond
$N=184$. Nevertheless some qualitative discrepancies between two MM
approaches can be found. In FRDM barriers are systematically higher by
over 2 MeV. Fission barriers of odd isotopes are also calculated and
discussed in the cited paper.

\subsection{Self-consistent theories}
\label{sec:scm}

The MM methods with their two independent ingredients (LD model for
bulk parameters and WS model for SP properties) lead to a rather large
number of theoretically independent parameters separately fitted in LD
and SP parts of the energy.
Contrary to this empirical treatment, self-consistent approaches start
from one density functional model for the energy of a nucleus as a
whole depending on a couple of densities (local density, kinetic
energy density, spin-orbit density, ...). These total energies can be
generated in the HF or HFB framework from given effective interactions
or modeled directly as density functionals. Such approaches exist in
both regimes, non-relativistic as well as relativistic.  In the
following we describe the HFB approach based on two most commonly used
(non-relativstic) functionals: the Skyrme type with zero-range nuclear
forces and the Gogny type where the forces are of mixed nature and
both zero-range and finite-range components are taken into account.

\subsubsection{Skyrme type functional}
\label{SSEC-stf}

The energy in Skyrme-Hartree-Fock (SHF) or
Skyrme-Hartre-Fock-Bogoliubov (SHFB) theories of nuclei reads
\begin{equation}
  \label{eq:3a:1}
  E = \int {\cal H}[\rho(\vec r),\tau(\vec r), \vec{J}(\vec r), \dots]dV\,,
\end{equation}
where $\rho$ is the local density, $\tau$ the kinetic energy density,
$\vec{J}$ the spin-orbit density, and so on.  (see {\it e.g.},
\cite{Vautherin72,Vautherin73}.)

There is a variety SHF energy density functionals which differ mainly
by their parametrisations.  One which is often used in the description
of the fission process is SkM$^*$ \cite{Bartel82}.  Heights of the
fission barriers known from experimental data and spurious rotational
energies in the ground states of even-even nuclei were taken into
account in the fitting procedure. Thus, fission properties should be
correctly calculated.

The rather involved calculations of barriers and the fission HLs
 of SH elements on a global scale were possible and done only since
very recently (see {e.g.}, \cite{Sch09a,Erl12a,Staszczak-13}).
The goal of the present section is to recalculate the results of
previous estimates for HLs taking into account two facts: First, as it
was mentioned in many places (see e.g.,~\cite{Warda2002,Libert1999}
where the rotational moment of inertia -- a part of inertia tensor --
is discussed), the Inglis cranking mass is too small in comparison to
the full ATDHFB one. In this paper, we use the cranking mass but
scaled with a factor 1.3 to simulate the ATDHFB mass. We follow this
principle in the present section. Second, a key ingredient is the
energy $E_0$ of the collective ground state; in earlier papers we used
rather crude estimates for $E_0$ and in the following we use an
improved semi-classical recipe (Bohr-Sommerfeld quantisation)
consistent with WKB approximation which is used to calculate the
tunnelling probability for spontaneous fission.  Note that other
calculations proceed diifferently in those details. For example,
reference~\cite{Sch09a} used fully quantum mechanically computed $E_0$
withint the Bohr Hamiltonian approach and the full ATDHF masses.
\begin{figure}
  \centerline{\includegraphics[width=.85\textwidth]{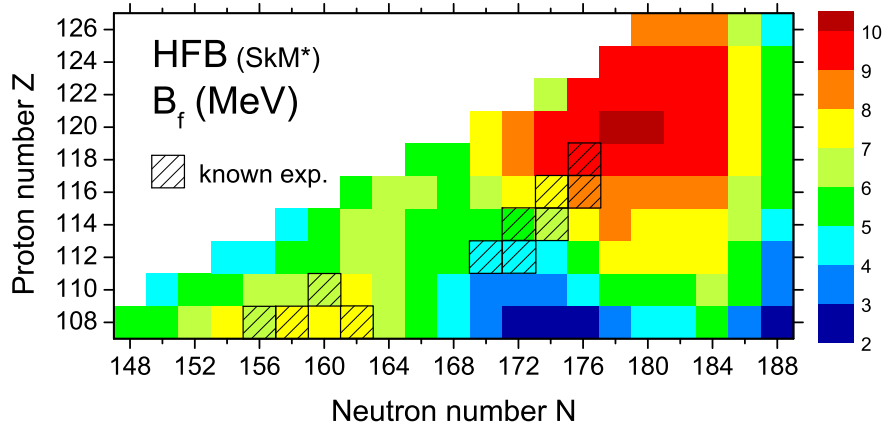}}
  \caption{\label{fig:3a:1} First fission barrier (MeV) of
    SH nuclei in HFB SkM$^*$ model. Cross-hatched squares represent
    observed nuclei.}
\end{figure}
Figure \ref{fig:3a:1} shows the systematics of the barrier heights
calculated with SkM$^*$ in the region of $Z=108(2)126$ and
$N=148(2)188$. In the experimentally known region, the barriers are of
the order of 6-7 MeV. The barriers of heaviest nuclei are relatively
high and reach 8-10 MeV. From the point of view of the barrier height
the most stable nuclei with respect to fission decay are close to
$Z=120-122$ and $N=178-180$. There is a region of low barriers height
near the line $A=282$. This would predict a valley of fission
instability in this particular region of superheavies. On the other
hand there is experimental evidence \cite{Oganessian12} which
contradicts calculations.
\begin{figure}[thb]
  \centerline{\includegraphics[width=.85\textwidth]{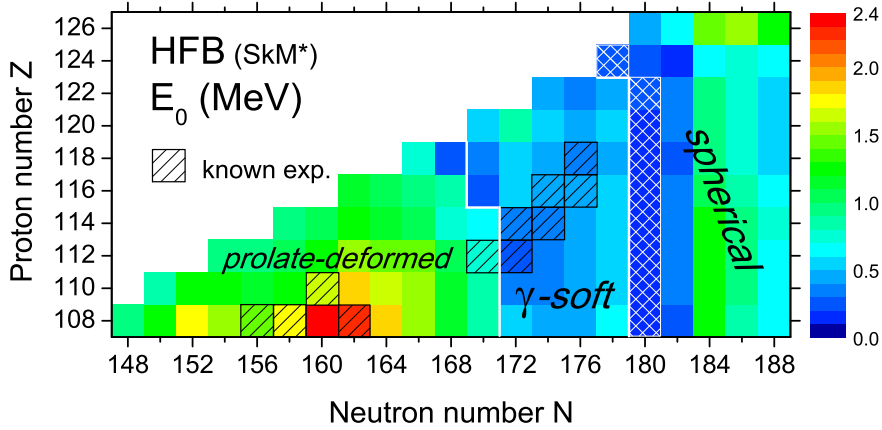}}
  \caption{\label{fig:E0} Ground state energies from WKB quantisation
    rule (see Eq.~\ref{eq:bsq}). Different deformation regions:
    prolate, $\gamma$-soft and spherical are approximately
    depicted. Cross-hatched squares represent observed nuclei.}
\end{figure}

The calculation of main ingredients of the theory, namely the PES and
the collective mass parameter were described
elsewhere~\cite{Baran11c,Staszczak-13}.  We map the deformation
landscape by adding a quadrupole constraint stretching the system
successively to scission.  The PES is then obtained as a function of
the mass quadrupole moment defined as the average value of the
quadrupole operator $q_2=\langle \hat{Q}_{20} \rangle = \langle 
\sum_{i=1}^{A}\left(2z_{i}^{2}- x_{i}^{2}-y_{i}^{2}\right) \rangle$.
Two further constraints, namely the triaxiality and reflection
asymmetry, were applied in the region of the first barrier and
beyond. A detailed discussion is given
in~\cite{Staszczak-13,Baran11c}. Here we only discuss some new aspects
concerning the spontaneous fission process.

The ground state energy measured relatively to the minimum of the
potential energy is crucial in determining the HLs of the nucleus with
respect to fission. The approximate fission probability $P$ stemming
from WKB theory involves mass parameters $B$, potential energy $V$ and
the energy of the collective ground state of the system $E_0$. The
requirement that the phases of WKB functions coincide (modulo $n\pi$)
in the region of their overlap leads to the quantization
condition. Then the following relation between the quantities involved
should be fulfilled for a stationary state~\cite{Schiff49}
\begin{equation}
  \label{eq:bsq}
  S_{ab} = (n+1/2)\pi\,,
\end{equation}
where, in general, the action $S_{ab}$ (in units of $\hbar$) calculated
between two points $a$ and $b$ is
\begin{equation}
  \label{eq-3-BS}
  S_{ab}(E_0) = \int_a^b \sqrt{2B(q)[E_0-V(q)]}dq\,.
\end{equation}
The coordinate $q$ is the leading collective coordinate towards fission
({\it e.g.}, $q=\langle\hat Q_{20}\rangle$ or $q=\beta$), $n=0,1,2$ {\it
  etc}. In Eq.~\ref{eq:bsq} $a$ and $b$ are the classical turning
points determined from the equation $V(q)=E_0$ solved in the vicinity
of the ground state. The Eq. (\ref{eq-3-BS}) is similar to the
Bohr-Sommerfeld quantisation rule (see e.g., \cite{Geldart+86}). It
can be easily solved for each considered nucleus with respect to
$E_0$.  While the fission probability $P$ strongly depends on $E_0$
the results of calculation will be different from those where one
assumes a constant $E_0=0.5$ MeV for all nuclei (see
e.g.~\cite{Baran05e,Warda2002}) or quantum mechanically exact values
\cite{Sch09a}. The values of the ground state energies $E_0$
calculated from the Eq.~\ref{eq:bsq} are shown in
Figure~\ref{fig:E0}. It is interesting to observe correllations of
this quantity with the corresponding ground state
deformations~\cite{Hee15}.

For the mass parameter we aim at the adiabatic time dependent HFB
(ATDHFB) cranking approximation.  As it was already mentioned, the
Inglis cranking mass parameters are on the average smaller then full
ATDHFB ones.  Therefore to do more realistic calculations of fission
HLs we assumed according to other authors the scaled fission mass
tensor component. The scaling factor 1.3 was applied. This leads in
general to larger values of $E_0$ [see Eq.~(\ref{eq-3-BS})] but HL
remains the same as mass parameters are larger.

The next Figure \ref{fig:3a:2} shows the fission HLs calculated in
this way with SkM* parameter set.
\begin{figure}[ht]
  \centerline{\includegraphics[width=.85\textwidth]{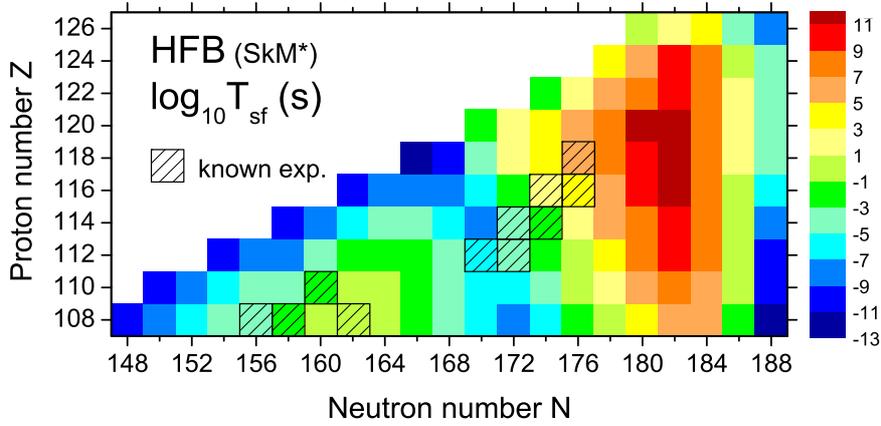}}
  \caption{\label{fig:3a:2} Fission HLs (sec) of superheavy nuclei in
    microscopic HFB SkM$^*$ model. Cross-hatched squares  represent
    observed nuclei.}
\end{figure}
\begin{figure}[ht]
  \centerline{\includegraphics[width=.85\textwidth]{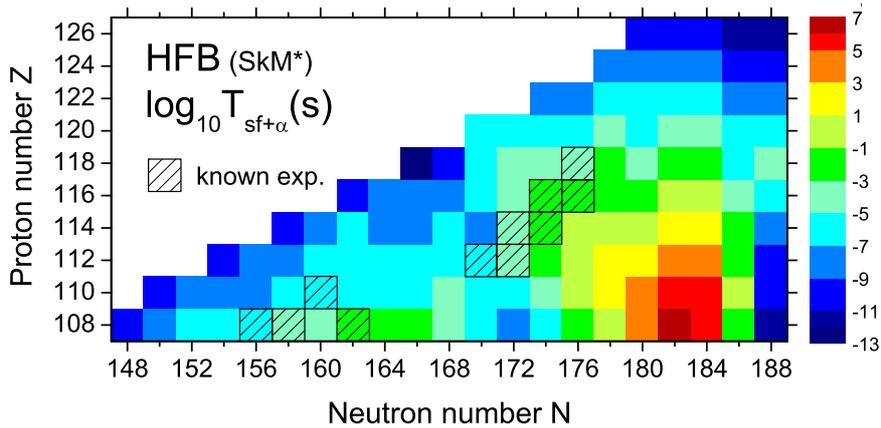}}
  \caption{\label{fig:3a:3} Total HLs (fission + $\alpha$) of
    superheavy nuclei in HFB SkM$^*$ model. Cross-hatched squares represent
    observed nuclei.}
\end{figure}
The WKB quantisation rule which delivers the ground state energy $E_0$
for fission works well at least in the region of known elements as
e.g., in the case of fermium isotopes \cite{Baran14b}.  Its
extrapolation to SH elements seems to be better than an {\it ad hoc}
procedure which was assumed previously where the ground state energy
$E_0$ was equal to $0.7\,E_{\mbox{ZPE}}$ where $E_{\mbox{ZPE}}$ is zero
point energy, or it is a constant equal to 0.5\,MeV.

After taking the $\alpha$-decay into considerations the HLs for some
nuclei change substantially.  To estimate $\alpha$-decay half-lives,
we used the standard Viola-Seaborg expression \cite{Viola66} with the
parameters from Ref.~\cite{Par05}.  The resulting HLs are shown in
Figure \ref{fig:3a:3}.
According to our calculations the longest total HLs correspond to
nuclei in the vicinity of $Z = 112$ and $N = 182,\,184$.  The region
of most stable nuclei is located in the vicinity of $Z\sim112$,
$N\sim182,\,184$~\cite{Staszczak-13}.
\begin{figure}[hbt]
  \centerline{\includegraphics[width=.85\textwidth]{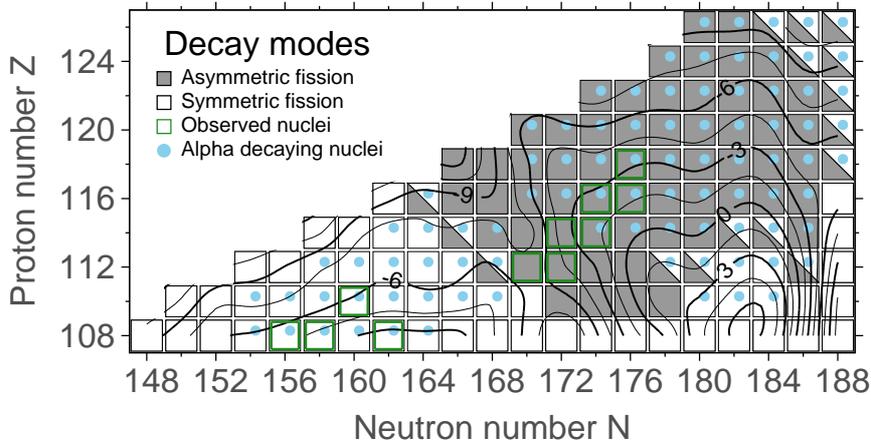}}
  \caption{\label{fig:3a:4} Regions of symmetrically and
    asymmetrically fissioning nuclei in HFB SkM$^*$ model. The filled
    circles (blue) represent nuclei decaying mainly by alpha
    emission. The isolines show approximate total HLs (in
    seconds). Experimentally registered nuclei are depicted as thick
    (green) squares.}
\end{figure}
It is worth to mention the existence of two main modes of fission.
The symmetric mode (the same spherical fragments) which dominates in
the region of light SH nuclei and the asymmetric one prevail in the
region of still heavier systems.
Both symmetry regions are shown in the Figure \ref{fig:3a:4}.  The
separation line is close to $A\sim280$. The asymmetric fission region
is concentrated at $Z\sim118$ and $N\sim178$. It is observed that the
heaviest nuclei ($N>186$) decays probably in the asymmetric way. The
thick squares shown in Figure~\ref{fig:3a:4} correspond to nuclei for
which experimental data for fission and/or alpha decay HLs are
known~\cite{Oganessian12}.  These trends are in agreement with the
survey of \cite{Erl12a} using a couple of other Skyrme
parametrisations.

Fission barrier heights and HLs with respect to spontaneous
fission as obtained from self-consistent HFB SkM$^*$ calculations
differ considerably from the ones presented in the previous section where
MM models were used. The main difference are the
barriers. They are usually higher in the case of HFB SkM$^*$ model up
to 2-3 MeV at some cases. This feature is common for all
non-relativistic self-consistent approaches.

There are of course discrepancies between the theoretical and the
experimental data, the largest ones in the region of the middle mass
SHE. Since those systems belong to the region of shape coexistence
some further increase of spontaneous fission HLs is anticipated due to
shape mixing. In order to improve upon the theory of fission we
shall consider the inclusion of several collective coordinates and the
dynamics of spontaneous fission process, the improved energy density
functionals~\cite{Kortelainen12}, and the full ATDHFB inertia
parameters.  The work is in progress. Some introductory results have
been already published~\cite{Jhilam13}.

\subsubsection{The Gogny functional}

The Gogny force \cite{Gogny.75} is a non-relativistic effective
phenomenological interaction widely used to describe the low energy
dynamics of the atomic nucleus. It is intended to be applicable to all
possible nuclei from drip line to drip line using a limited set of
parameters adjusted to global properties. In addition to the standard
central, spin-orbit and Coulomb potentials it contains a
phenomenological density dependent term depending on the density
raised to the power $\gamma=1/3$ introduced to incorporate the
saturation mechanism of nuclear forces. In many aspects, it is similar
to the Skyrme \cite{Skyrme56} interaction sharing with it the
spin-orbit potential, Coulomb and the idea of a density dependent
repulsive part. However, Skyrme's zero range and gradient terms
present in the central potential are replaced in the Gogny force by a
finite range potential linear combination of two gaussians. The main
advantage of the finite range is that the corresponding matrix
elements are free from ultraviolet divergences allowing the treatment
of the particle-particle (pairing) channel with the same interaction.

In most of the fission studies carried out with the Gogny force, the
focus has been in the characterisation of just one of the fission
observables, namely the spontaneous fission lifetimes (or derived
quantities like fission barrier heights and widths) in even-even
nuclei. Studies aimed at the description of induced fission in the
framework of finite temperature HFB are scarce \cite{Martin.09} and
deserve more attention. This is the same as with fission in odd mass
nuclei where the understanding of the specialisation energy is of
fundamental interest. In this respect, there is only one example with
the Gogny force \cite{Perez.09}. Recent advances in the experimental
techniques allow for the study of the evolution of fission parameters
with spin \cite{Henning.13}.  The expected decrease of the barrier
height with spin and the evolution of triaxiality in the first barrier
has been studied with the Gogny force in Ref \cite{Egido.00}.

At this point, it has to be mentioned that the ability of any of the
Gogny parametrisations to reproduce fission observables (mainly
spontaneous fission HLs) is spoiled by the strong dependence of
those quantities with the value of the collective inertia in the
forbidden region.  As the collective inertia strongly depends on the
pairing gap as the inverse of its square it is to be expected that
pairing properties should play a central role in the theoretical
description of fission. In this respect, we can mention two studies
\cite{Guzman14,Guzman14b} with D1S, D1N and D1M where the spontaneous
fission lifetimes of the neutron rich uranium and plutonium isotopes
have been considered. The variation of the HLs with the
pairing strength was found to be enormous in absolute value although
the trend with neutron number was roughly preserved.  As a consequence
of this variability, the results also depend upon the principles
involved in the determination of the path to fission. The recent results
of \cite{Giuliani.14} using the minimum action principle with pairing
degrees of freedom are markedly different from those using the minimum
energy principle ({\em see} Section \ref{sec:dyn}). The reason again is
the strong dependence of the collective inertia on the pairing gap
(it goes as the inverse of the square of the pairing gap) that has a
strong impact on the action.

The PES calculated in a microscopic model like HFB theory provide a
lot of information on the fission properties of SH elements. HLs,
preferred decay channels and fission types (symmetric, asymmetric,
highly asymmetric) can be deduced by the analysis of PESes. As it was
mentioned above, results derived within HFB theory are obtained with
some systematic theoretical errors ({\it see\/}
Section~\ref{sec:err}). Nevertheless, it is reasonable to compare
results between particular isotopes and distinguish groups of nuclides
with specific decay properties. An early study of SH region in the HFB
model has been performed by Bruy\`eres-le-Ch\^atel group
in \cite{Decharge1999, Berger:2001aa, BHG-2003, dbgd2003, BHGD-2004}
and Lublin-Madrid group in \cite{WER-06}. The most thorough
investigation using the D1S Gogny interaction in the axial symmetry
regime has been published in
\cite{Warda:2012aa}. Details of the method are described in
\cite{Warda2002}.

\begin{figure}
  \centerline{\includegraphics[width=0.9\columnwidth]{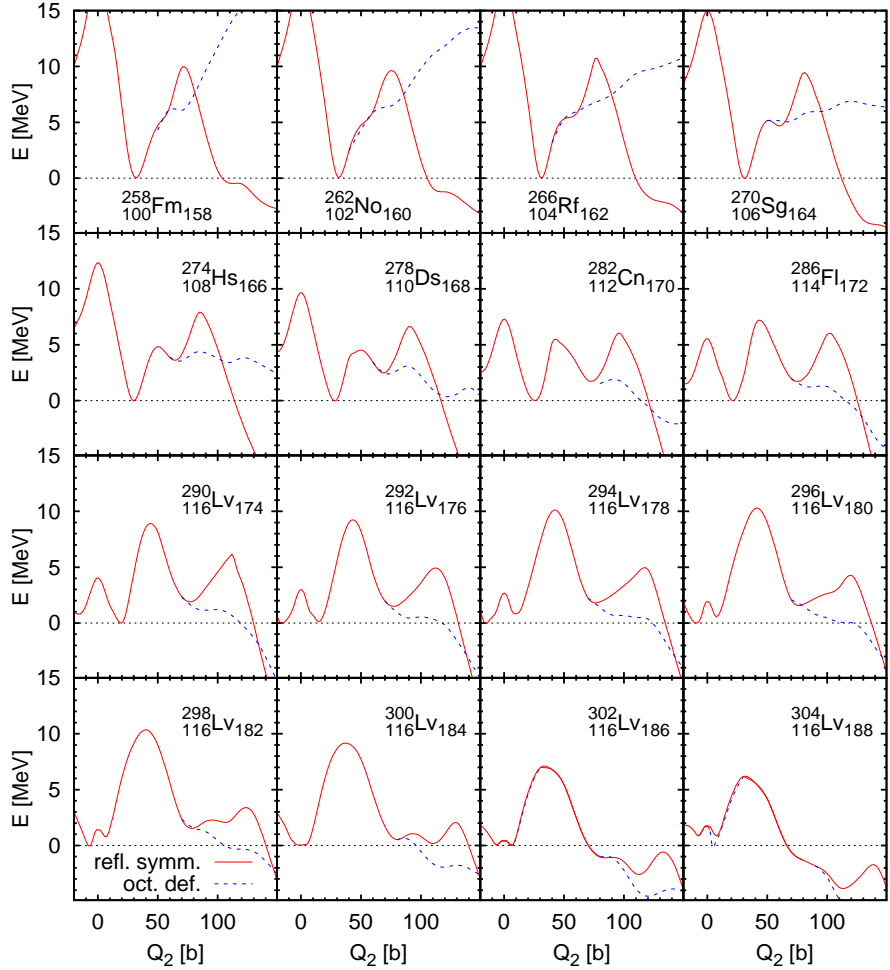}}
  \caption{\label{fig:3b:1} Axial fission barriers of selected
    isotopes across the SH nuclei. Note that in the original paper
    \cite{Warda:2012aa} the quadrupole moment $Q_2$ was a factor of
    two smaller.}
\end{figure}
Fission characteristics are determined not only by the height of the
fission barrier but also by its shape and localisation in the
deformation space.  From the general overview of the fission barrier
shapes across the SH nuclei  one can find
that the number of protons affects mainly the height of the barrier whereas
substantial changes of the barrier profile take place along isotopic
chains. In consequence, fission properties depend mostly on neutron
number.  Selected fission barriers of SH isotopes calculated in the
axial regime are presented in Fig. \ref{fig:3b:1}.

\begin{figure}
 \centerline{ \includegraphics[angle=270,width=0.9\columnwidth]{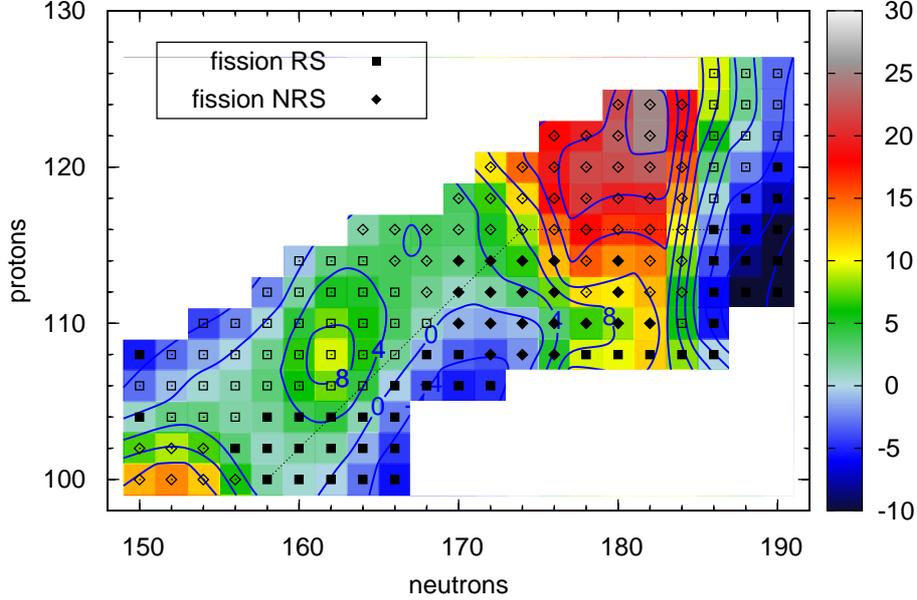}} \caption{\label{fig:3b:2}
  Logarithm of the spontaneous fission HLs in seconds computed with
  the HFB model and the D1S Gogny force. Nuclides with dominant
  fission are represented by filled symbols whereas those with
  dominant $\alpha$ decay are marked by open symbols. Dotted line
  indicates isotopes which fission barrier is presented in
  Fig. \ref{fig:3b:1}.}
\end{figure}
The lightest SH elements, with neutron number below $N=170$, are well
prolate deformed in the ground state. The saddle point in these nuclei
has very high energy, often exceeding 8 MeV.  Going along any isotopic
chain in this region from $N=150$ to $N=162$ one can observe
increasing width of the fission barrier while the barrier height
remains almost unchanged.  As a consequence, the action integral
across the barrier increases and longer HLs are observed (see
Fig. \ref{fig:3b:2}).  The centre of the region of extended stability
is predicted for $^{270}$Hs with the longest fission HL
$\log(t_{\mathrm{SF}}/\mathrm{s})=12.83$. The corresponding nucleon
numbers, namely $N=162$ and $Z=108$, are sometimes called ``deformed
magic numbers''. The special characteristics of the $N=162$ and
$Z=108$ system are correlated with the decrease of the neutron pairing
energy in the ground state and a small value of the proton one
\cite{Warda:2012aa}.  This is consistent with the aforementioned MM
and Skyrme model results.

In most of the isotopes below $N=162$, the fission barrier goes
through reflection symmetric shapes. Only in the nuclei in the
transitional region between the actinides and the SH elements with
$A\sim250$ octupole shapes play an important role in the determination
of the fission barrier height. The reflection symmetric character of
the barriers brings about symmetric fragment mass distribution in this
region. The barrier height may be reduced by a few MeV in the whole
region if non-axial deformations are included in the calculations
\cite{Warda2002}.

Beyond $N=158$, we find several different trends in the structure of
the PES.  First, the top of the fission barrier splits into two peaks:
the first one is lower and the second one higher. In isotopes heavier
than $N=162$ the height of both saddles decreases and, in consequence,
HLs of subsequent isotopes become shorter.  The second, reflection
symmetric saddle may be additionally diminished if other degrees of
freedom characterising the nuclear shape are allowed. Octupole
deformation cuts off the second barrier, but up to $A=280$ such
reflection asymmetric barriers extends to large quadrupole
deformations and can not become the dominant fission channel. In
heavier nuclides the width of the octupole deformed hump in the
fission path is comparable to the reflection symmetric one and
asymmetric fission can be observed as a dominant fission channel. Also
triaxial shapes can decrease the height of the second barrier but at
the cost of increasing the inertia and therefore the contribution to
the action remains unchanged~\cite{Warda:2012aa}.
   
In the same range of neutron numbers, the ground state deformation
decreases and energy difference between prolate and oblate minimum
reduces. In isotopes with $178\le N\le182$ the oblate minimum becomes
the ground state.  When non-axial deformations are taken into account,
the local minimum with higher energy becomes a saddle point and there
is a path along the $\gamma$ plane connecting it with the ground
state.

As a consequence of the aforementioned changes of the PESs, fission
HLs decrease with the mass of nuclides.  Minimal values, not longer
than $t_{\mathrm{SF}}=10^3$~s, are reached around neutron number
$N=170$ in each SH element.

The first peak of the barrier keeps the axial reflection symmetric
character in the whole region above $N=158$. Its height shows a
growing trend with both neutron and proton number in isotopes between
$N=170$ and $N=182$ which is responsible for increasing fission HLs up
to $N=182$.  The isotopes with two neutrons more ($N=184$) have got
features of magic nuclei: well pronounced spherical ground state with
small pairing energy and large shell gap in neutron single-particle
energy spectrum. However, a lower fission barrier induces here larger
fission probability than in lighter isotopes. There is no strong
indication of magic character of any proton number in this region of
SH elements in the calculations with the Gogny force.

Beyond $N=184$ the second, octupole deformed barrier drops away below
the ground state energy level and fission barrier becomes
substantially narrower. This change dramatically reduces fission HLs
by several orders of magnitude with adding subsequent pair of
neutrons. On the other hand, in the nuclei with neutron number greater
than $N=184$ octupole deformation decreases the ground state energy by
a varying amount that can reach the 2 MeV. This decreasing of the
ground state energy work in the direction of increasing again the
height of the fission barrier. However, this effect is not strong
enough as to counteract the above mentioned reduction of the fission
HLs.

Two regions of extended stability against fission can be found in the
SH elements, see Fig. \ref{fig:3b:2}. The first is around the deformed
magic nucleus $^{270}$Hs where fission HLs reach the level of
$10^{12}$ s. The second region covers isotopes with neutron number
$N=182$. The HLs in elements with large proton number reach here much
larger values extending far above $t_{\mathrm{SF}}=10^{20}$ s. It is
important to notice the lower stability against fission in isotopes
around $N=170$ which is presently in the focus of experimental
investigations. In nuclei heavier than $N=186$, a decrease of
potential energy with deformation dictated by LD bulk properties
dominates over the stabilising shell correction. High fission barrier
can thus not be achieved and the fission stability drops rapidly.

The main decay process competing with spontaneous fission in SH
elements is $\alpha$ radioactivity. The predicted $\alpha$ decay HLs
are shorter than fission ones in proton rich isotones. Thus, it can be
noticed that fission is the dominant decay channel in three regions
\cite{Warda:2012aa}. First, in the vicinity of deformed magic nucleus
$^{270}$Hs symmetric fission can be found. Second, asymmetric fission
is dominant for nuclei elements lighter then Fl ($Z=114$) and below
magic neutron number $N=184$. Finally, for heavy isotopes with
$N\ge186$ and $Z\le118$ again symmetric fission occurs with very short
HLs.

In those isotopes beyond $N=190$, the macroscopic part of the nuclear
potential makes fission barriers too small to allow for sufficiently
long HLs that could be detected experimentally. The existence of
heavier nuclei is possible only in exotic ground state
configurations. Bubble nuclei, spherical symmetric systems, without
nuclear matter in the centre, are predicted for charges $Z\ge 240$ and
masses $A\ge 700$ \cite{Decharge1999, Berger:2001aa, BHG-2003,
  dbgd2003, BHGD-2004}. Lighter nuclei with $Z\ge 170$ and $A\ge 450$
can exist in semi-bubble configuration where central density is
reduced to half of the bulk nuclear matter density.  Another type of
exotic shape that may exist beyond SH nuclei is the one of a torus
\cite{Warda-07}. The HFB theory with Gogny force predicts a toroidal
structure with energy lower than spherical configuration for nuclei
with $Z>140$. The stability of such exotic shapes is questionable and
it has to be confirmed in further investigations.

\subsubsection{Relativistic Mean Field}

Fission barriers in SH nuclei has been also investigated in covariant
density functional theory \cite{Abu2012}. Modern parametrisation
DD-ME2 and DD-PC1 was used together with modified version of
traditional NL3*. It was found that both triaxiality and octupole
deformation substantially affect the second barrier in nuclei with
$Z\ge112$. It was found that the first barrier remains axially
symmetric in the whole region.

Triaxial deformation decrease energy of the saddle point as much as 3
MeV in some isotopes with $N<174$. The NL3* and DD-ME2 predict the
lowest saddle point energies among the all models of nuclear
forces. The DD-PC1 parametrisation gives slightly larger values.

\subsection{Mass parameters and their impact on half-live times}
\label{sec:mass}

The collective mass along the fission path influences the fission
HLs with the same importance as the barrier does.  This can be seen from 
formula (\ref{eq-3-BS}) for the action integral where the mass $B(q)$
and the potential energy $V(q)$ enter the action as a product. A
relatively recent discussion of the mass tensor calculations can be
found in~\cite{Baran11c}.

Mass parameters are usually calculated within perturbative cranking
approximation (or Inglis inertia) or adiabatic time dependent HF or HFB
model where one differentiates the fields of the model numerically
instead ( see, {\em e.g.},
\cite{Goe87aR,Libert1999,Warda2002,Sch09a,Giannoni80,Giannoni80a}).

The shapes of the mass tensor components are fairly similar in both
mentioned cases but the values of specific components differ
substantially. The simple cranking masses calculated in {\em e.g.},
one dimensional space are usually smaller then ATDHF cranking masses.
This influences, in turn, the fission HLs. In the EDF of the
Gogny type models of nuclear structure where the barriers are much
higher then ones from {\em e.g.}, MM or RMF models, these small masses
lead to smaller HLs, which sometimes coincide with experimental
ones. On the other hand, the Skyrme EDF give smaller barriers then
Gogny ones and the ATDHF cranking masses are numerically
calculated. They generate HLs much larger or comparable to that of the
Gogny model. In the MM models the barriers are believed to fit rather
well experimental data and one uses the simple perturbative cranking
approximation for the masses.  Unfortunately that leads to too small
(in comparison with experiment) HL times. The other feature is
the collective inertias strongly depend upon the pairing strength
what makes the pairing degree of freedom an important ingredient
(see the section on dynamics of fission (\ref{sec:dyn})).

The above discussion shows that there is still a multitude of
different approaches to compute the collective mass along the fission
path. This hinders a direct comparison of the final fission HL. It is
a challenge for the future projects to remove all of these
differences in the calculations of the nuclear inertia by adopting
the same ATDHF cranking model everywhere. Only then it will be
possible to judge the models and the validity of the barrier heights
as well as the validity of final HL.

\section{\label{sec:compar}Comparison of calculated fission barriers
  to experimental data}
There are two main methods of calculating fission HLs and
barriers, namely the static method (minimisation of the potential
energy and extraction of the saddle points) and dynamic (minimisation
of the action integral).  Most of the calculations (and these in this
paper) concerning fission barriers are of the static type.

\subsection{Static fission barriers}
\begin{figure}[hbt]
\centerline{\includegraphics[width=.8\textwidth]{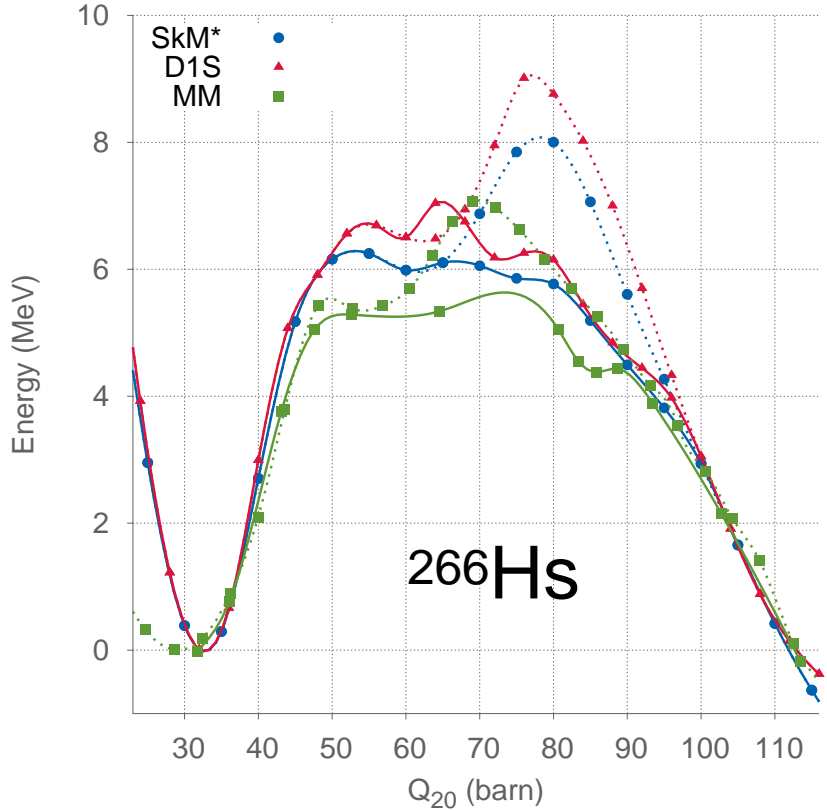}}
\caption{\label{fig:bar-hs}Fission barriers of $^{266}$Hs as functions
  of the quadrupole moment $Q_{20}$ calculated in MM model (green,
  filled squares), Skyrme HFB approach (blue, filled circles) and
  Gogny HF model (red, filled triangles).  Dashed lines represent the
  raw axial barrier.  Solid lines correspond to the barriers with
  $\gamma$. The values of $Q_{20}$ for MM curves are close to the
  actual, but not completely accurate.}
\end{figure}
We start with the comparison of the fission barriers for $^{266}$Hs as
an example of mid-mass SH isotope calculated in models discussed in
the paper. The barriers derived in the MM, Skyrme HFB and Gogny HF
models are shown in Fig.~\ref{fig:bar-hs} as functions of quadrupole
moment $Q_{20}$.  The shapes of all PES are very similar and there is
relatively good agreement of barrier heights. The differences in the
peak height (the saddle point) reach 1 MeV.  Rather small differences
remaining are hard to explain because the models are very different in
their structure. The differences shown will weakly impact on
calculated fission HLs.
\begin{figure}
  \centerline{\includegraphics[width=.49\textwidth]{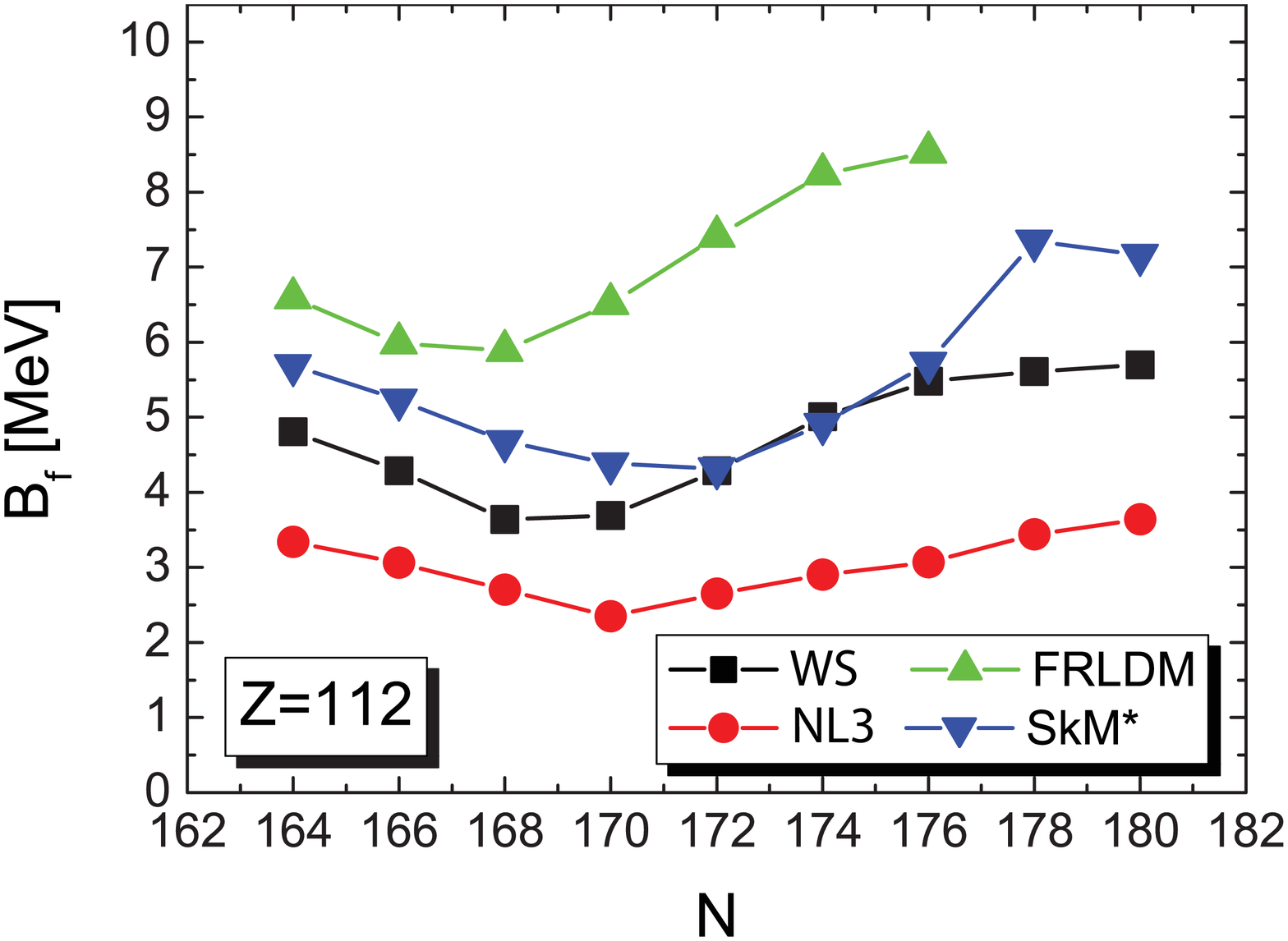}
    \includegraphics[width=.49\textwidth]{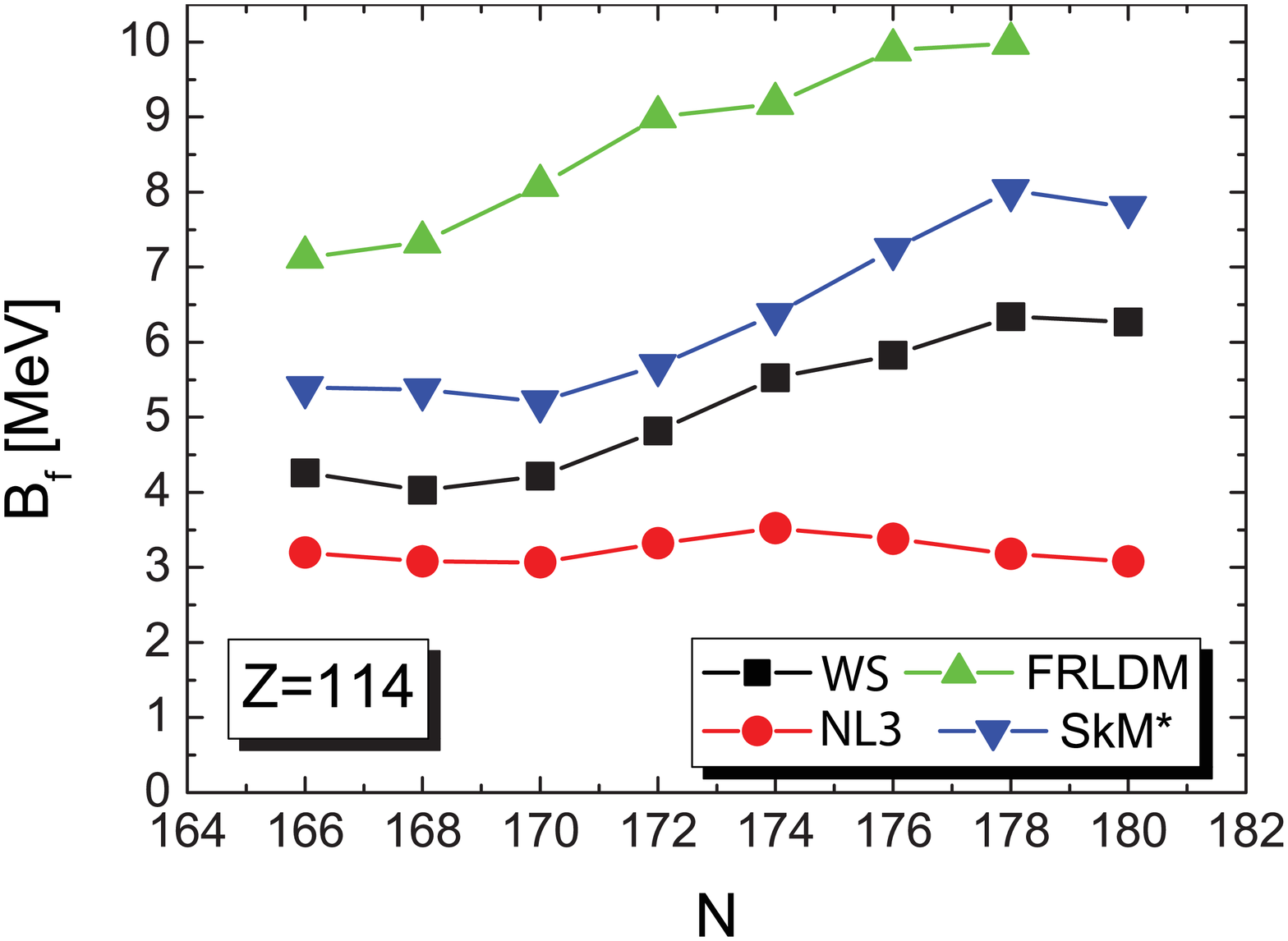}}
  \centerline{\includegraphics[width=.49\textwidth]{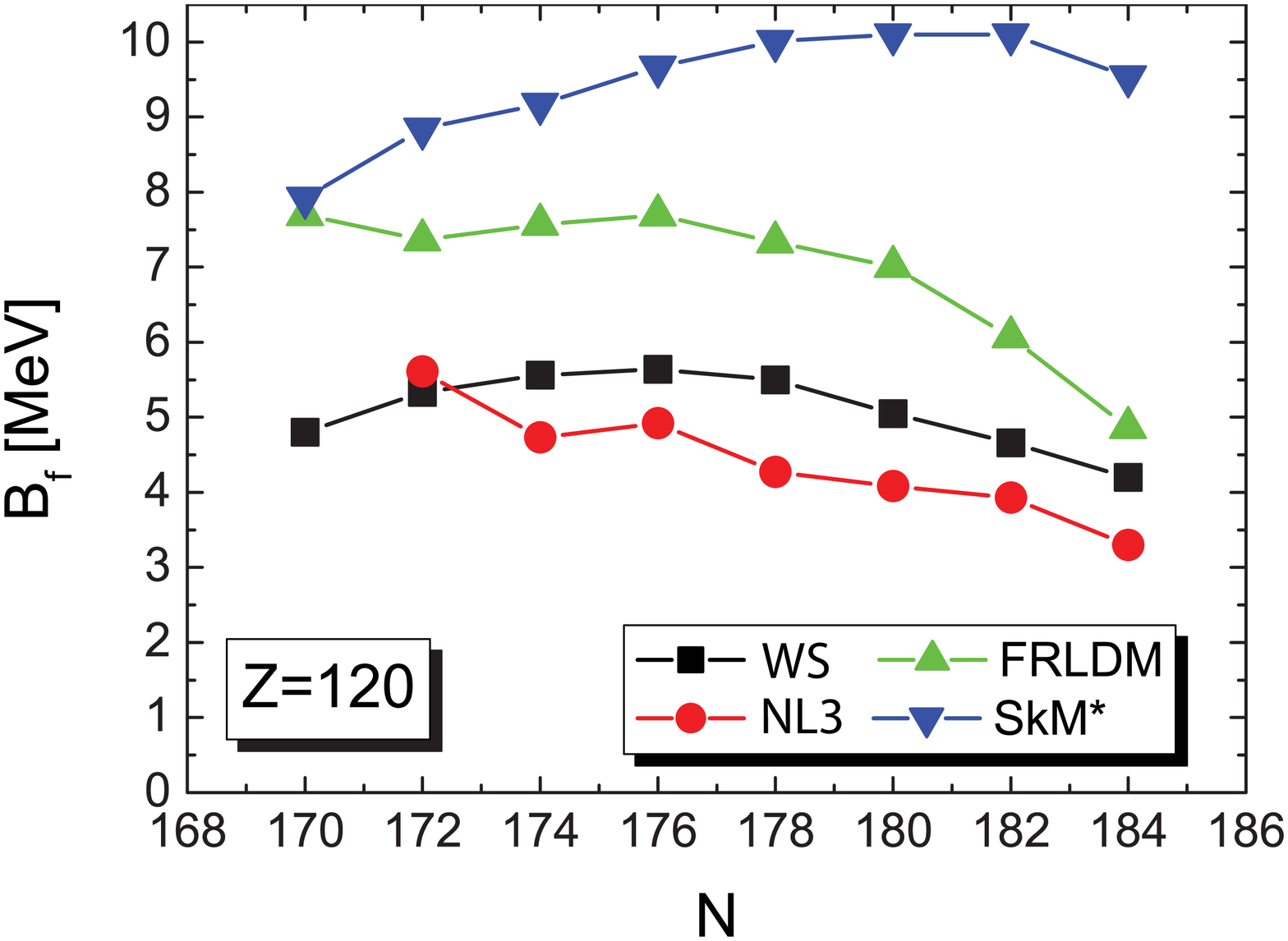}}
  \caption{{\protect Fission barriers predicted by various models:
      black - WS model~\cite{Kowal2010a}, green -
      FRLDM~\cite{Moller2009}, blue - SkM*~\cite{Staszczak-13}, red -
      RMF with NL3 parametrisation~\cite{Abu2012}.}}
  \label{fig:bar}
\end{figure}
Fig. \ref{fig:bar} shows four sets of calculated fission barriers
$B_f$ for even-even isotopes of $Z=112, 114$ and 120: from the WS
model~\cite{Kowal2010a}, the macroscopic-microscopic Finite Range
Liquid Drop Model (FRLDM)~\cite{Moller2009}, HFB calculations with the
SkM* interaction~\cite{Staszczak-13} and the RMF model with NL3
parametrisation 
\cite{Abu2012}. Results of rather comprehensive calculations were
selected: all were made for dozens of nuclei and included important
triaxial and mass-asymmetric deformations. One has to emphasise that
three of those models well reproduce experimental barriers in
even-even actinides; the SkM* model overestimates actinide barriers by
1-2 MeV. In spite of this, the differences in predictions in
Fig. \ref{fig:bar} are conspicuous.  In order to quantify this, we
refer to the known spontaneous fission (s.f.) HLs: 0.8 ms in
$^{282}$Ds, 97 ms in $^{284}$Ds and 130 ms in $^{286}$Fl. The values
predicted some time ago~\cite{SSS} from the sub-barrier action
minimisation based on the WS model were, respectively, 71 ms, 4 s and
1.5 s (1-2 orders of magnitude overestimate; only axial deformations
were included in~\cite{SSS}, but axially symmetric shapes at the
barrier in these nuclei were actually confirmed in~\cite{Kowal2010a}).
One can observe that the FRLDM barriers for $Z=112, 114$ are much
higher than in the WS model and could hardly produce HLs in agreement
with experiment.
As we have checked, this huge difference in barriers in the two MM
models is not related to shape parametrisation.  The spontaneous
fission HLs for the three nuclei calculated in~\cite{Staszczak-13}
along static path (no action minimisation) are (surprisingly, in view
of barriers similar to ours) by many orders of magnitude too short.
On the other hand, RMF calculations~\cite{Abu2012} produced barriers
much lower than in the WS model which could hardly explain the
measured evaporation residue cross-sections in fusion.

In Table \ref{5} we collect some of the theoretical predictions of
fission barrier heights based on the FRLDM~\cite{Moller2009}, SHF --
self consistent Hartre-Fock method~\cite{BUR2004} with the SLy6 Skyrme
interaction~\cite{CHAB98} and the Extended Thomas-Fermi plus
Strutinsky Integral model (ETFSI) based on Skyrme SkSC4 functional
\cite{MAM2001} in relation to experimental ones~\cite{ITKIS001} and
our predictions in a MM and SHFB(SkM*) models. Note that the lower
limits for the fission barrier heights are evaluated in
Ref.~\cite{ITKIS001}. As we can see, experimental and MM calculated
barrier heights are in agreement while FRLDM, SHF and SHFB generally
overestimate the barrier heights.  Surprisingly the modern version of
the MM calculation presented in Ref.~\cite{Moller2009} (including
nonaxiall shapes) gives values comparable to axially-symmetrical
calculations of SHF and overestimate the experimental barriers in a
similar way. This tendency is visible, to a lesser extent, also for
heavier systems.  Thus for the isotope $^{292}$Ds$_{176}$ the FRLDM
model gives a value of 9.25 MeV for the barrier height while the value
obtained in our approach (6.22 MeV) is about 3 MeV lower, whereas the
experimental data indicate 6.4 MeV. For the neighbouring isotope
$^{292}$Lv$_{178}$ the experimentally predicted fission barrier is also
6.4 MeV and we obtain also almost the same value 6.28 MeV, but
estimations based on the FRLDM predict a value of 9.46 MeV, almost 3
MeV larger. As a consequence, such high barriers result in cross
sections for Z=114,116,118 and 120 which overestimate the experimental
data by several orders of magnitude~\cite{SIWEK09,SIWEK10}. The
ETFSI(SkSC4) model significantly underestimates the barrier heights
for $^{284}$Cn$_{172}$ and $^{286}$Cn$_{174}$ for heavier systems the
agreement is however much better. One must keep in mind that
ETFSI(SkSC4) and SHF(SLy6) models do not include triaxial nor
higher order nonaxial degrees of freedom and the observed
discrepancies probably can be explained by the absence of these
variables at the saddle point configuration.
\begin{table}
\begin{center}
  \caption{\label{table:bar}Comparison of fission barrier heights with
    another theoretical evaluations: SHF(SLy6)~\cite{BUR2004},
    SHFB(SkM*) -- this paper, FRLDM~\cite{Moller2009}, ETFSI(SkSC4)
    with Skyrme SkSC4 force~\cite{MAM2001}, WS~\cite{Kowal2010a} --
    present paper, and experimental data taken
    from~\cite{ITKIS001}. \label{5}}
  \medskip
\begin{tabular}{ccccccc}
\toprule[1pt]
\addlinespace[.3em]
   Nucleus  &   SHF   &  SHFB   & FRLDM &  ETFSI  & WS & EXP  \\
            &  (SLy6) &  (SkM*) &       & (SkSC4) &    &   \\
\midrule[.2pt]
   $^{284}112_{172}$  & 6.06 & 4.31 & 7.41 & 2.2 & 4.29 &   5.5   \\
   $^{286}112_{174}$  & 6.91 & 4.74 & 8.24 & 3.6 & 5.01 &   5.5   \\
   $^{288}114_{174}$  & 8.12 & 6.24 & 9.18 & 6.1 & 5.53 &   6.7   \\
   $^{290}114_{176}$  & 8.52 & 7.13 & 9.89 & 6.6 & 5.83 &   6.7   \\
   $^{292}114_{178}$  &  -   & 8.14 & 9.98 & 7.2 & 6.34 &   6.7   \\
   $^{292}116_{176}$  & 9.35 & 8.72 & 9.26 & 6.5 & 6.22 &   6.4   \\
   $^{294}116_{178}$  & 9.59 & 8.84 & 9.46 & 7.2 & 6.28 &   6.4   \\
   $^{296}116_{180}$  &  -   & 8.54 & 9.10 & 7.2 & 6.07 &   6.4    \\
   $^{294}118_{176}$  &  -   & 9.19 & 8.48 & 6.6 & 5.99 &    -   \\
   $^{296}118_{178}$  &  -   & 9.47 & 8.36 & 7.0 & 6.04 &    -   \\
   $^{298}118_{180}$  &  -   & 9.28 & 8.05 & 7.4 & 5.72 &    -   \\
   $^{296}120_{176}$  &  -   & 9.48 & 7.69 & 6.2 & 5.64 &    -   \\
   $^{298}120_{178}$  &  -   & 10.05 & 7.33 & 6.6 & 5.50 &    -   \\
   $^{300}120_{180}$  &  -   & 9.91 & 7.01 & 6.8 & 5.05 &    -   \\
   $^{302}120_{182}$  &  -   & 9.82 & 6.07 & 7.2 & 4.66 &    -   \\
   $^{304}120_{184}$  &  -   & 9.54 & 4.86 & 6.8 & 4.20 &    -   \\
\bottomrule[1pt]
 \end{tabular}
\end{center}
\end{table}

\subsection{\label{sec:dyn}Fission dynamics}
As it was shown years ago~\cite{Baran78,Baran81} and also in recent
publications~\cite{Jhilam13,Jhilam14,Guzman14,Kaska14} the
minimisation of the reduced action integral instead of the potential
energy leads to larger fission probabilities or equivalently to
shorter fission HLs. The latter are generally reduced by one to two
orders of magnitude in comparison to statically calculated HLs.  At
the same time, the calculated fission barriers increase by 1-2
MeV. While these kinds of calculations are rather involved, one
usually performs static calculations of HLs integrating the action
along the potential energy valley leading to the fission of the
nucleus. This is common to all calculations.

The dynamic treatment of some internal degrees of freedom like {\em
  e.g.}, pairing gap ({\it see\/} \cite{Pilat89} and the very recent
paper~\cite{Giuliani.14}) or particle number fluctuation
parameter~$\lambda_2$ in Lipkin-Nogami pairing
model~\cite{Guzman14,Jhilam13,Jhilam14} seems to be more significant
than a similar treatment of space-type deformation coordinates
described above. While the latter is already well known to be similar
for all instances the former depends on the considered nucleus.  For a
few cases, it was shown in the above cited papers that a dynamical
treatment of the pairing field is important and one should include
pairing fluctuations as a dynamical variable in the calculation of
HLs.  The dynamical change of {\em e.g.}, the pairing gap being the
consequence of action minimisation corresponds to the fission path on
which the collective mass parameter is generally smaller than one on
the static pathway and at the same time the fission barrier increases
as compared to the static one.  The final HL is reduced again. The
reduction factor is around one order of magnitude. For reasons of
space, we do not discuss this type of calculations in detail here.

\section{\label{sec:err}Uncertainties on barriers and correlations with
other observables}

So far, we have mostly presented the results for the various models
separately. A direct comparison of barriers for four selected models
was done in Figure \ref{fig:bar} and it shows that the predictions on
fission properties can vary widely amongst the models. It is the aim
of this section to explore the variations of predictions on a broader
and more systematic basis.  To this end, we will exploit statistical
analysis built on least-squares techniques for optimising
parametrisation.  We will briefly summarise the least-squares
techniques, then discuss the various influences on barriers by
explicit variation of force parameters, and finally present
systematics of extrapolation uncertainties and correlations with
system properties from statistical analysis.  The basic principles of
$\chi^2$ error estimates and correlations analysis have been explained
in great detail in \cite{Dob14a}.  A practical application in the
context of SHF with inspecting a broad range of observables is found
in \cite{Erl14a}.  Following the scope if this paper, we concentrate
here on fission properties.  As we have seen in figure \ref{fig:bar}
that the isotopic and isotonic trends of fission properties are about
the same for the different models, we confine the study to one
representative test case, fission properties of $^{266}$Hs as a
prototype super-heavy element.

\subsection{The fission path of $^{266}${\em Hs}}

The fission path is a set of mean-field states $|\Phi_{q}\rangle$
comprising a range of relevant quadrupole deformations $q$.  In
self-consistent mean-field models, it is generated by mean-field
calculations with quadrupole constraint, see {\it
  e.g.}~\cite{Goe87aR}.  This yields the potential-energy surface
(PES) for fission.
\begin{figure}[ht]
\centerline{
\includegraphics[width=0.8\linewidth]{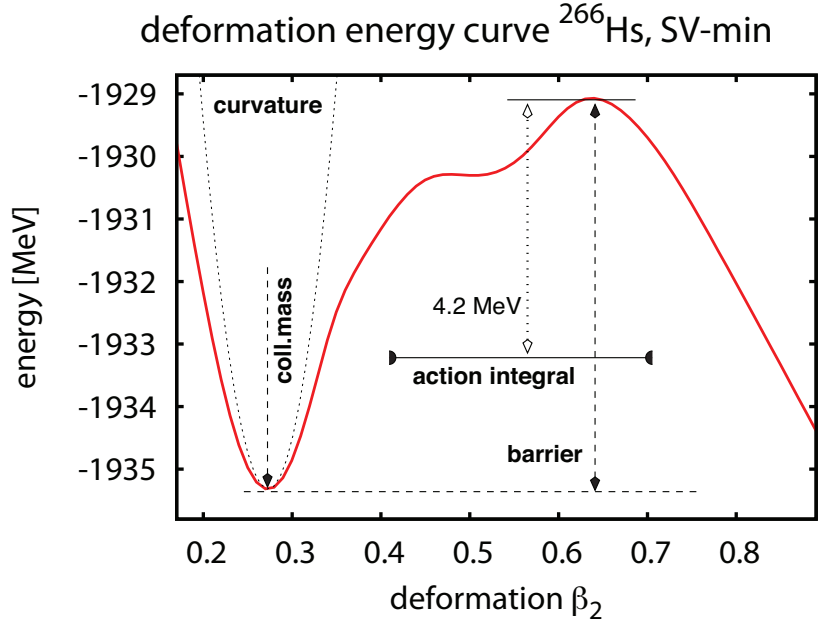}
}
\caption{\label{fig:PEStypical-2} Illustration of the potential-energy
  surface (PES) along the axially symmetric fission path of $^{266}$Hs
  and its four key quantities: curvature of the PES at the ground
  state, fission barrier, mass parameter $B$ at the ground state, and
  the action integral
  $=\int_{0.4}^{0.7}d\beta_2\sqrt{B(V_\mathrm{PES}-E_\mathrm{sub})}$
  with $E_\mathrm{sub}=V_\mathrm{max}-4.2$MeV as appropriate average
  quantities in the tunnelling regime. The PES was computed with
  quadrupole constrained SHF calculations using the parametrisation
  SV-min.}
\end{figure}
Figure \ref{fig:PEStypical-2} shows the PES for our test case
$^{266}$Hs.  It is what we call the ``raw PES'' which emerges as
expectation value of energy for given deformation. The actual fission
calculations will invoke also the mass parameter along the path and
augments the raw PES with quantum corrections (zero-point energies)
\cite{Goe87aR,Sch09a}. This modifies the PES and with it the
barriers. Fortunately, the modifications are moderate for $^{266}$Hs.
This allows us to confine the discussion of trends and correlations to
properties of the raw PES. For simplicity, we will also confine the
considerations to the fission path along axially symmetric shapes.  It
is known that triaxiality produces for $^{266}$Hs typically about 1 - 2
MeV lower barriers (cf. Sec. \ref{sec:compar}.1 and Fig. \ref{fig:bar-hs}).  
However, axial shapes suffice for the purpose of
comparing different parametrisation and studying trends and
correlations.

For a systematic analysis, it is advantageous to reduce the
information on the PES to a few key quantities.  Figure
\ref{fig:PEStypical-2} illustrates the choice of these key
observables. The most obvious quantity is the height of the fission
barrier $B_f$. In the following we consider $B_f$ as a difference
between the saddle point of PES and the energy of mean-field ground
state. (Note that the effective barrier seen in experiment is rather
the difference between the saddle point energy and the collective
ground-state energy which lies typically 1-2 MeV above the mean-field
ground state energy.)  The collective ground state is characterised by
the curvature
$\left.\partial^2E(q)/\partial{q}^2\right|_{q_\mathrm{min}}$ and the
mass parameter (not shown here) at the minimum.  Another key piece for
determining fission HL is the WKB action integral (see
Eq.~\ref{eq-3-BS}) for tunnelling. This is a very involved quantity as
it depends on entry and exit point for the tunnelling regime which, in
turn, depend sensitively on the energy of the collective ground
state. We want here a simpler measure concentrating on the mass
parameter in the tunnelling regime. As this parameter varies
considerably along the path (see figure \ref{fig:path}) we consider an
average taken over the core of the tunnelling regime.  To that end, we
consider the action integral $S$ (see caption) in the form as it
enters the tunnelling integral. In order to concentrate on the effect
of the mass $B(q)$, we confine the integration regime safely inside
the turning points and take a fixed energy of 4.2 MeV. Of course, the
actual value for the action integral depends on this reference
energy. But we are interested in comparing differences which develop
when varying the forces. The trends seen in such variations are the
robust against the detailed choice of reference energy and integration
limits.

\begin{figure}[ht]
\centerline{
  \includegraphics[width=0.7\linewidth]{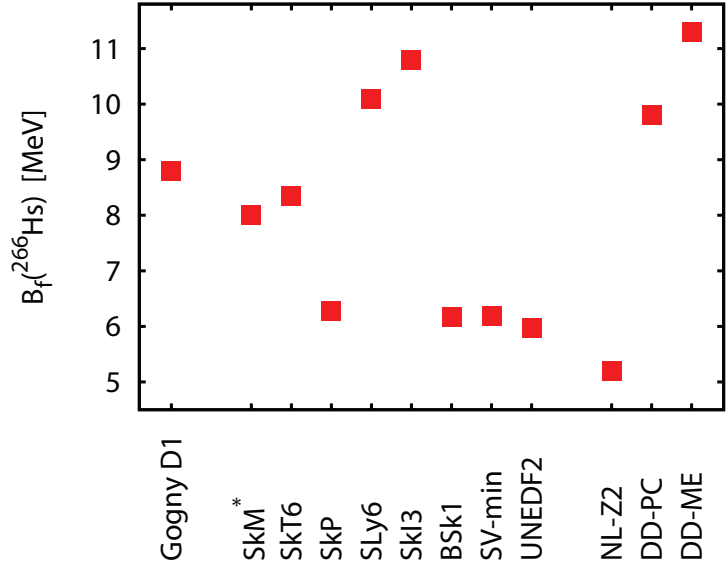}
}
\caption{\label{fig:collect-barriers-266Hs} The axially symmetric
  fission barrier of $^{266}$Hs for a great variety of published
  models and parametrisation.  The basic citations for the models and
  parametrisation are: Gogny \cite{Dec80a},SkM$^*$ \cite{Bartel82},
  SkP \cite{Dob84a}, SkT6 \cite{(Ton84)}, SLy6 \cite{Cha97a}, SkI3
  \cite{Rei95a}, BSk1 \cite{Sam02a}, SV-min \cite{(Klu09)}, UNEDF2
  \cite{Kor14a}, NL-Z2\cite{Ben99a}, DD-PC and DD-ME \cite{Naz14c}.  }
\end{figure}
Before proceeding to $\chi^2$ analysis, we illustrate in figure
\ref{fig:collect-barriers-266Hs} the broad span of predictions for the
most important quantity of the fission path, the fission barrier of
$^{266}$Hs. To that end, we have fetched a couple of published
parametrisation from early first generation models to latest
developments. There are three formally different models involved, the
Gogny force to the left, a large group from SHF in the middle, and
three RMF models at the right. Even within one model family, SHF or
RMF, one sees a large variation of results and it is obvious from this
figure that many of the parametrisation have to be ruled out for
fission calculations.  For SHF we can establish a trend of $B_f$ with
the effective mass $m^*/m$ of a parametrisation where low $m^*/m$ is
related to large $B_f$ and vice versa. This argument does not apply to
the three RMF parametrisation which all have about the same low
$m^*/m$. Here it is a strongly varying incompressibility $K$ which
causes the different $B_f$. After all, the example of figure
\ref{fig:collect-barriers-266Hs} demonstrates that an arbitrary
collection of parametrisations does not allow to conclude
unambiguously on the relations between properties of a force and
predicted observables (here $B_f$). Statistical analysis based on
$\chi^2$ fits provides a well defined strategy to disentangle the
various influences and to estimate uncertainties in predictions. This
is what we will unfold now in this section. Before carrying on, we
want to remark that statistical analysis can unravel trends,
uncertainties, and correlations for a given model. Besides that we
have to be aware of a systematic error which sneaks in by the choice
of a functional form for the mean-field model. These can be explored
by variation of a model and comparison of different models
\cite{Dob14a} as exemplified in \cite{Erl14a}. We will not pursue the
quest for systematic errors here.

\subsection{Brief review of $\chi^2$ analysis}
\label{sec:chi2}

The Skyrme-Hartree-Fock approximation is a typical
representative of presently used self-consistent nuclear models.  Its
structure can be deduced from general arguments of low-momentum
expansion \cite{(Neg72),Rei94aR} while the remaining model parameters
are determined by adjustment to empirical data, for reviews see
\cite{(Ben03),Sto07aR}.  Strategies for such adjustment have much
developed over the decades.  Present standard are least-squares
($\chi^2$) fits started in \cite{Fri86a} and steadily extended to
include more and more data, for recent examples see, {\it e.g.},
\cite{(Klu09),(Kor10)}. 
The least-squares ($\chi^2$) fits and subsequent analysis of
extrapolation errors as well as correlations are a standard technique
and well documented at several places
\cite{Dob14a,(Bra97a),(Bevington)}.  The tasks is to optimise a set of
model parameters $\bld{p}=(p_1,...,p_{N_p})$ such that a
representative set of observables $\{\hat{\mathcal{O}}_i,i=1...N_d\}$
is optimally reproduced. 
This is achieved by defining a quality function $\chi^2(\bld{p})$
from the sum of squared errors and minimising it.
Extrapolation uncertainties and correlations between observables can
be estimated from the behaviour of $\chi^2$ in the vicinity of the
minimum. To that end, one defines probability distribution
of ``reasonable'' parametrisations as 
$W(\bld{p})\propto\exp\left(-\frac{1}{2}\chi^2(\bld{p})\right)$.
The (statistical) expectation value of an observable $\hat{A}$ is then
$\overline{A}=\int d^{N_p}p\,A(\bld{p})\,W(\bld{p})$.
Extrapolation uncertainties quantify the fluctuation around the mean
value and are computed as
$\Delta{A}=\sqrt{\overline{\left(A-\overline{A}\right)^2}}$.
The correlation between
$A$ and $B$, also called covariance, becomes
$c_{AB}=\overline{\left(A-\overline{A}\right)}\overline{\left(B-\overline{B}\right)}/(\Delta{A}\Delta{B})$.
The actual evaluation is usually done in terms of Gaussian integrals
because $\chi^2$ can be well approximated by a quadratic form in
$\bld{p}$, see {\it e.g.}~\cite{Dob14a}.

\subsection{Choice of fit observables and force parameters}

For the fit observables entering the $\chi^2$, we consider here the
same set as it was used in the survey \cite{(Klu09)}: binding
energies, key parameters of the charge form factor (r.m.s. radius,
diffraction radius, surface thickness), odd-even staggering of binding
energies, spin-orbit splittings in semi-magic nuclei selected to be
well described by a pure mean-field model. A straightforward fit to
these ground state data yields the parametrisation SV-min. It embraces
all information how the given ground state data determine features of
the parametrisation and how much they leave undetermined.

A key feature of a model are the nuclear matter properties (NMP).  We
will consider here the four NMP which characterise the response
properties of nuclear matter: incompressibility $K$, isoscalar
effective mass $m^*/m$, symmetry energy $a_\mathrm{sym}$, and TRK
sum-rule enhancement $\kappa_\mathrm{TRK}$ (for a detailed definition
see {\it e.g.},~\cite{(Klu09)}). These NMP have been used since long as
benchmarks in various nuclear models and thus they have developed an
intuitive meaning. Moreover, there is a one to one relation between
NMP and the force parameters of the volume part of the SHF
functional. Thus NMP are equivalent to force parameters.  Dedicated
variation of NMP is useful tool to explore influences within a
model. Thus we will also consider parametrisations which are fitted as
SV-min, but with an additional constraint on NMP.  There is, for
example, the parametrisation SV-bas which was fitted while freezing $K$,
$m^*/m$, $a_\mathrm{sym}$, and $\kappa_\mathrm{TRK}$ at values such
that one obtains additionally also a good description of giant
resonance excitations \cite{(Klu09)}.  The difference between SV-bas
and SV-min is one indicator for the influence the these four NMP.  We
will consider in section \ref{sec:trends} an even greater variation of
parametrisations.

\subsection{Extrapolation uncertainties}

\begin{figure}[ht]
\centerline{
\includegraphics[width=0.4\linewidth]{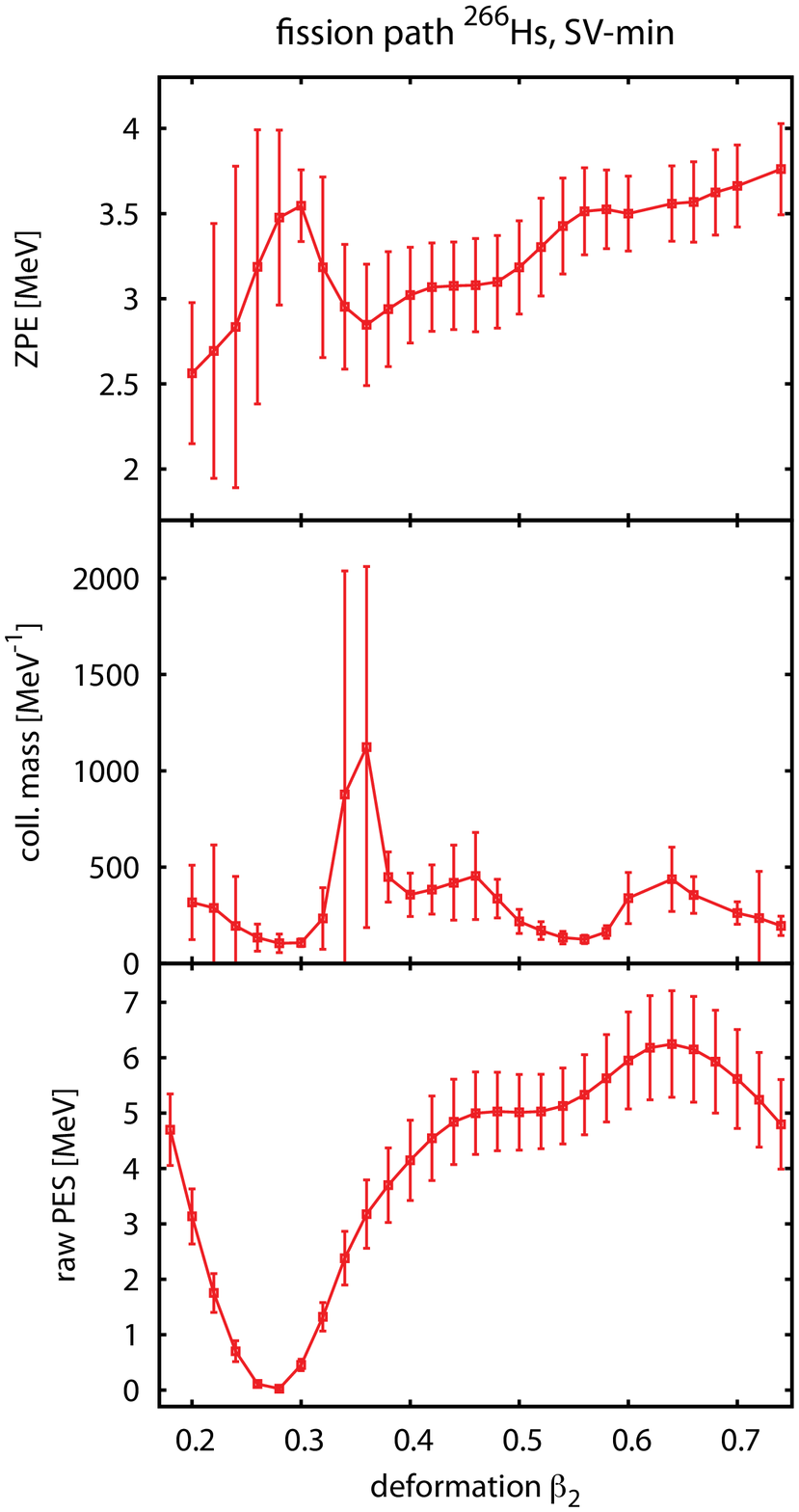}
$\;$
\includegraphics[width=0.4\linewidth]{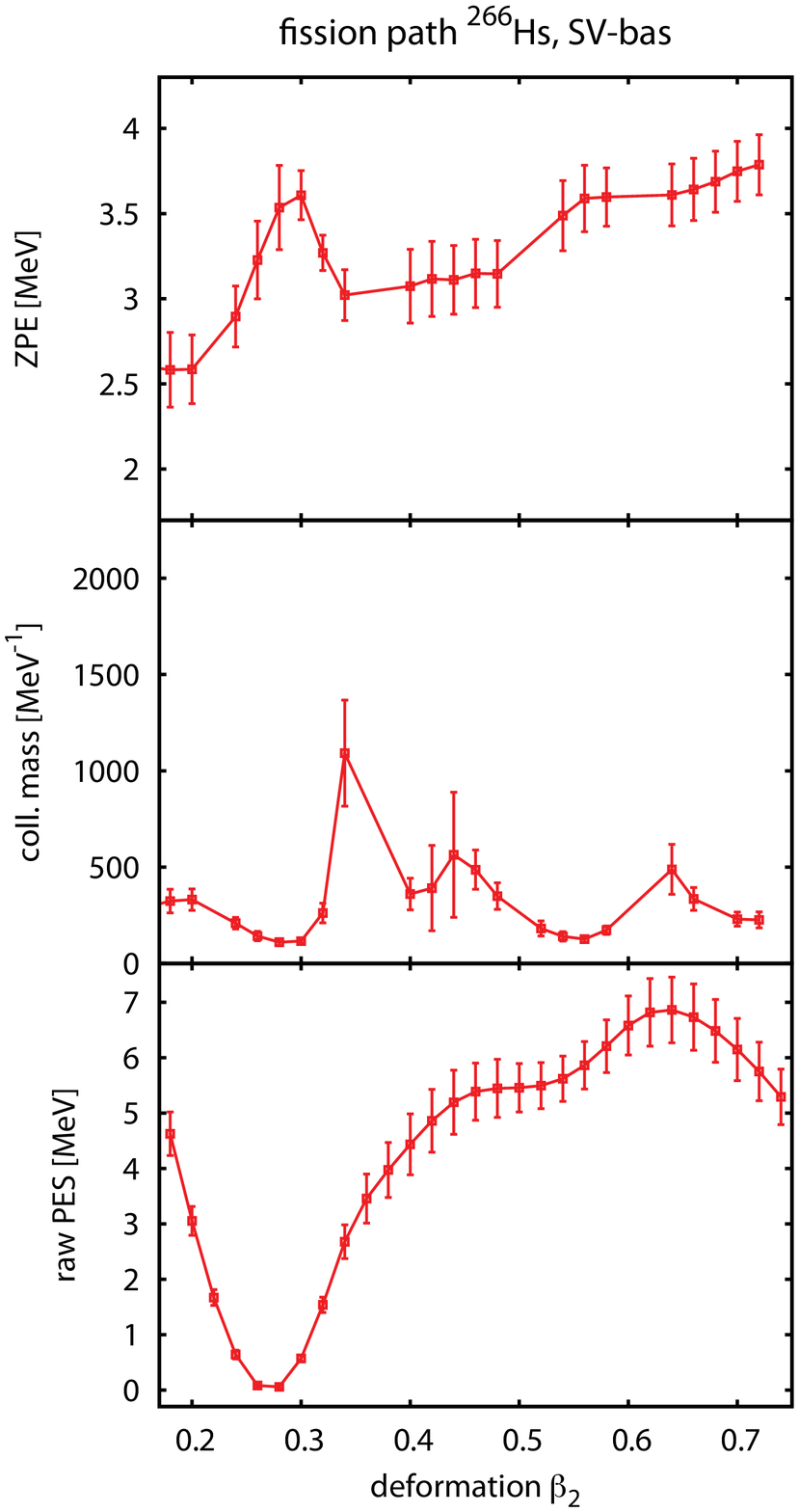}
}
\caption{\label{fig:path} Raw PES (lower panels), mass parameter
  (middle panels), and ZPE (upper panels) along the axial and
  reflection symmetric fission path of $^{266}$Hs. The left panels
  show results for SV-min, the right ones for SV-bas. Each case is
  shown together with the extrapolation uncertainties from $\chi^2$
  analysis.  }
\end{figure}
Uncertainties for extrapolated observables are the simpler signal from
$\chi^2$ analysis. Thus we start with a quick glance at uncertainties
for the present test case, fission of $^{266}$Hs. Figure
\ref{fig:path} shows the ingredients for a fission calculation,
potential-energy surface (PES), zero-point energy (ZPE) and mass
parameter, all for motion along axially symmetric quadrupole
deformation expressed as the dimension-less $q\equiv\beta_2$ (for
details of the collective picture behind see, {\it
  e.g.},~\cite{Goe87aR,Erl12a}).  The figure label indicates ``raw
PES''. This is the energy directly from a quadrupole constraint SHF
calculation. The collective fission dynamics employs the PES corrected
by ZPE \cite{Goe87aR,Erl12a}. The pattern for the corrected PES are
very similar to those for the raw PES and thus we have skipped this
extra panel.  The mass parameter is obtained by self-consistent
cranking, also called ATDHF cranking \cite{Goe87aR,Erl12a}.
All three quantities are shown together with the extrapolation
uncertainties from $\chi^2$ analysis (see section \ref{sec:chi2})
drawn as error bars.  Note that the PES are scaled to zero energy at
the minimum. There is thus no uncertainty at this point and only
slowly growing uncertainty around it. Sufficiently far away from the
minimum, the error bars are of similar size along the path. The same
holds for the ZPE.  More fluctuations of uncertainties are seen for
the the mass parameter. Some points have very large uncertainties and
some points for SV-bas did not produce a reasonable result at all
(that is why there are missing points in the curves). These artifacts
are due to round-off problems in the very demanding error analysis
combined with the demanding computation of mass parameters. These
problems arise typically in regions where many level crossing exist.

The uncertainties for the unconstrained fit SV-min are larger than
those for SV-bas.  This is plausible because SV-bas is constrained by
four NMP which reduces uncertainties.  However, the reduction is only
about factor two. The constraint on $K$, $m^*/m$, $a_\mathrm{sym}$,
$\kappa_\mathrm{TRK}$ thus has only partially impact on fission
properties. There remains an equally strong influence from the other
force parameters.
The uncertainty of the PES at the barrier is 1 MeV for SV-min and 0.6
MeV for SV-bas. This shows the benefit of adding more information from
NMP, alias giant resonances \cite{(Klu09)}. Still the uncertainty is
large as compared to the extreme sensitivity of the fission
lifetime. A prediction within four order of magnitudes is already great
success \cite{Erl14a}.

\subsection{Correlation analysis for the key observables}
\label{sec:correlkey}
In a study of inter-correlations within the fission path (not shown
here) we have found that the values of potential energy $V(\beta_2)$
along the fission path are highly correlated with each other which
means, {\it e.g.}, that knowing the barrier height $B_f$ determines
the fission potential $V(\beta_2)$ in a broad vicinity around the
barrier.  The same hold for the mass parameter.  This allows to reduce
the relevant information to the few key observables as indicated in
figure \ref{fig:PEStypical-2}. We will henceforth reduce the
discussion to these.
\begin{figure}[ht]
\centerline{
\includegraphics[width=0.9\linewidth]{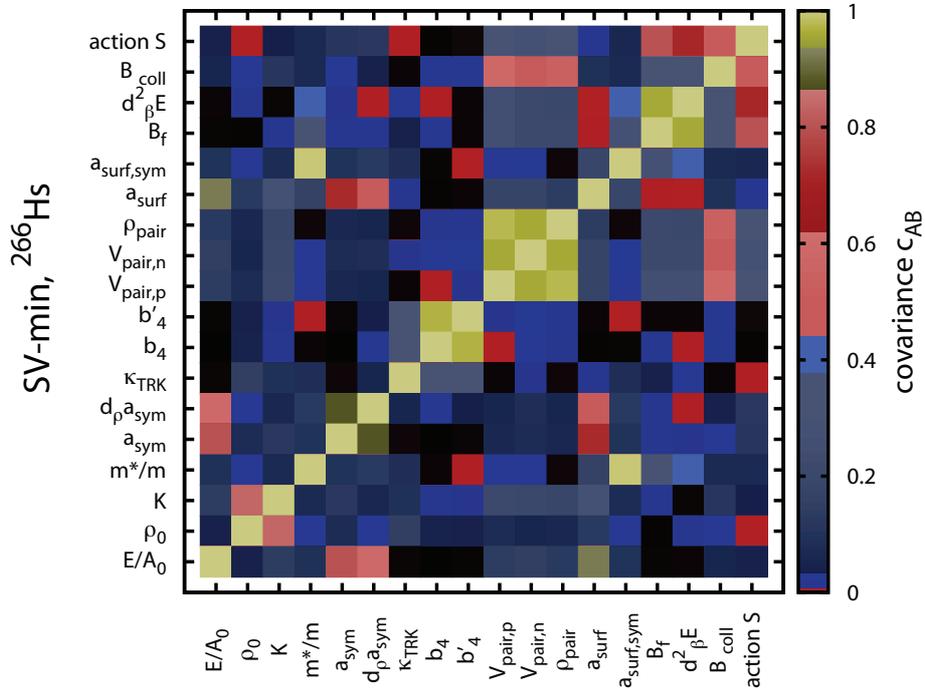}
}
\caption{\label{fig:alignmatrix} Matrix of correlations between the
  key observables of the fission path in $^{266}$Hs for selected force
  parameters SV-min.  }
\end{figure}
Figure \ref{fig:alignmatrix} shows the matrix of correlations between
forces parameters and key observables of the fission path for SV-min.
Remind that many of the SHF force parameters can be mapped to NMP
which are a more intuitive way to characterise a force. The five bulk
parameters of the SHF functional (volume term, density dependent term)
can be represented by the five NMP equilibrium binding $E/A$ with
$\rho_0$, incompressibility $K$, symmetry energy $a_\mathrm{sym}$ and
slope thereof $\partial_\rho a_\mathrm{sym}$. The two parameters for
kinetic terms correspond to the two NMP for effective mass, $m^*/m$
and $\kappa_\mathrm{TRK}$. The two parameters for the gradient terms
correspond to the two surface energies $a_\mathrm{surf}$ and
$a_\mathrm{surf,sym}$ computed here in semi-classical approximation
\cite{Sto81a}. Only pairing and spin-orbit parameters have no bulk
equivalent and remain what they are.

Amongst the key observables of the path (see figure
\ref{fig:PEStypical-2}), we see a strong correlation between barrier
$B_f$ and curvature
$\left.\partial^2E(q)/\partial{q}^2\right|_{q_\mathrm{min}}$, some
correlation between these two quantities and the action integral, and
very little correlation to the mass parameter at ground state.  What
the correlations of fission properties with force parameters is
concerned, they are mostly small. Only $m^*/m$ and pairing parameters
show some correlations, pairing particularly with the mass parameter
and $m^*/m$ with the barrier.
After all, we see that there is no pronounced correlation of a fission
observables to any particular force parameter.  This indicates that
each detail of the force contributes a bit to fission where from
pairing is the strongest player.

\subsection{Trend analysis for the key observables of $^{266}${\em Hs}}
\label{sec:trends}

Covariance analysis is a powerful tool to find out dependencies
between observables (and force parameters). But it often lacks
intuitive insight. As a complementing tool, one can perform a trend
analysis. This is done by producing systematically various sets of
forces where in each set one forces parameter (or NMP) is varied. Such
sets had been produced in \cite{(Klu09)} starting with SV-bas as base
point. The sets were obtained by constraining the same four NMP as for
SV-bas while varying one of the four NMP with respect to SV-bas. One
can then draw the observables of interest versus the varied
parameter. This makes the dependencies graphically visible which is
often a helpful complementing illustration of the correlation.

\begin{figure}[ht]
\centerline{
\begin{tabular}{l|cccccc}
\hline
       & $K$  & $m^*/m$ &$a_\mathrm{sym}$
       & $\kappa_\mathrm{TRK}$ & $V_\mathrm{prot}$
       & $V_\mathrm{neut}$ \\
       &  [MeV] &  & [MeV]  &  & [MeV$\,$fm$^{-3}$]   &  [MeV$\,$fm$^{-3}$]\\
\hline
 param.   & 233.5 & 0.9 & 30 & 0.4 & 675 & 607
\\
 $\Delta$ param. &8.5 & 0.07 & 1.9 & 0.3 & 25 & 39 
\\
\hline
\end{tabular}
}
\medskip
\centerline{
\includegraphics[width=0.85\linewidth]{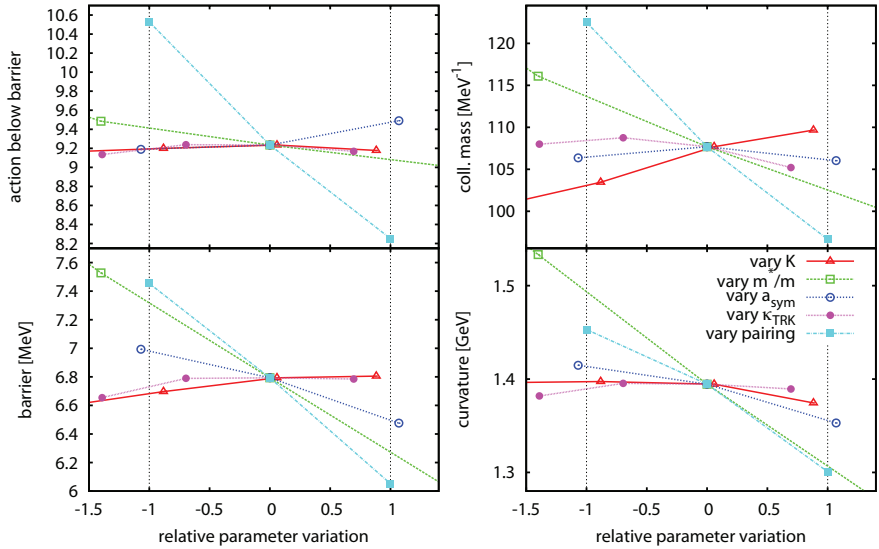}
}
\caption{\label{fig:trends-keypath} Trends of the key observables of
  the fission path of $^{266}$Hs with systematically varied force
  parameters.  The parameters are drawn relative to their reference
  value from SV-bas and scaled to the extrapolation uncertainty of
  SV-min. These reference values and their uncertainties are given in
  the table above the plots.  The values $+1$ for the scaled
  parameters stands for the variance shown in the table. The band $\pm
  1$ for the scaled parameters represents the allowed error band from
  $\chi^2$ analysis.}
\end{figure}
Figure \ref{fig:trends-keypath} illustrates the influences on fission
in terms of the trends with the force parameters (NMP, respectively)
$K$, $m^*/m$, $a_\mathrm{sym}$, $\kappa_\mathrm{TRK}$, and pairing
strength. In order to compare the trends within one panel (for each
observable), we produced dimensionless force parameters. This is done
by taking the parameters relative to the SV-bas point and scaled by
the uncertainty from SV-min which characterises the allowed variation,
{\it i.e.} the rescaled version of a parameter $p$ is
$p_\mathrm{rescal}=(p-p_\mathrm{SV-bas})/\Delta{p}_\mathrm{SV-min}$.
This implies that the interval of rescaled force parameters
$[-\!1,+\!1]$ represents the range of reasonable parameters for which
the reproduction of data remains nearly as good as for the minimum
($\chi^2$ growing at most by one unit
\cite{(Bra97a),(Bevington),Dob14a}).

The trends in Figure \ref{fig:trends-keypath} confirm what we have
seen already from the correlation matrix.  Pairing strength has the
strongest impact on fission. Next important is $m^*/m$.  There is also
some impact from $a_\mathrm{sym}$, but $K$ as well as
$\kappa_\mathrm{TRK}$ play little role here in SHF.

\section{\label{sec:summary}Summary and outlook}

We have presented a variety of theoretical results on fission
properties in super-heavy nuclei. Rather different modelling has been
compared, the macroscopic-microscopic (MM) method, one the one hand,
and self-consistent mean-field models on the other. The basis of the
description is the same in both model families: the state of the
system is characterised by a set of single-particle (SP) wavefunctions
moving in a common mean-field and the fission path consists in a
series of such states with successively increasing deformation until
two separated fragments emerge. The SP wavefunctions serve to account
properly for the quantum shell effects which are crucial for binding
of heavy elements and the topology of the potential energy
surface. The fact that a couple of different mean-field states is
involved means that the description deals, in fact, with collectively
correlated states. These features are the same for all
models. Different is the way in which the mean field and the total
energy are defined.  The MM method models the mean-field directly as a
(deformed) shell-model potential and composes the total energy from
the macroscopic liquid-drop model plus shell corrections deduced from
the s.p. energies. Self-consistent models start from an energy-density
functional where-from the mean-field equations are derived
variationally. The dedicated deformation which are needed to span the
fission path are obtained by adding a (quadrupole) constraint.  The
density functionals manage to incorporate a correct description of
bulk properties (liquid-drop parameters) together with shell
structure.  Within the self-consistent family, we have considered
three different models, the Skyrme-Hartree-Fock (SHF) approach which
is a non-relativistic density-functional method, the Gogny force
which, again, uses non-relativistic kinematics but a finite-range
force with a sprinkle of density dependence, and the relativistic
mean-field model (RMF) which relies on a covariant Lagrangian density
functional.

In spite of the very different modelling, the topological properties of
the fission path are coming out the same way in all models: ground
state deformations, fission isomers, shapes at barrier (reflection
symmetry, triaxiality), islands of stability. These features are
determined by quantum shell structure which is obviously correctly
reproduced in all cases. Huge differences in the predictions are found
in the quantitative aspects, height of fission barriers, and even more
so for fission lifetimes. These values are very hard to control. Each
detail of a model has an influence. Each one of the liquid-drop bulk
properties has some impact on barriers, the effective mass which
regulates shell effects plays also a role, and last not least pairing
is a crucial ingredient. Statistical analysis on top of least-squares
fits within the SHF approach estimates the uncertainties on barriers
typically of the order of $\pm 1$ MeV. We have to add substantial
leeway for systematic errors which, however, cannot be estimated in
systematic manner. The situation is even more involved for fission
lifetimes. Further ingredients are required for its computation: entry
point for spontaneous fission, collective mass along the path, and
quantum corrections for spurious zero-point energies. All these are
extremely subtle quantities which depend very sensitively on all
details of a model and on the way they are computed.  There is lot of
development work yet to be done to reach better agreement in the
quantitative predictions. For the time being, the method of choice is
to calibrate barriers and lifetimes for the heaviest known nuclei to
allow safer predictions for the even heavier elements.

Most self-consistent calculations describe fission as a
one-dimensional, adiabatic process. MM calculations indicate that this
may not always be sufficient. There are situations in which
deformation valley and fission valley go in parallel for while. It is
likely that one needs here a description in terms of multi-dimensional
collective motion. This is an extremely challenging problem which has
not been attacked thoroughly yet. We hope that increasing computing
power allows steps into that direction. The quality of the adiabatic
approximation deserves also critical testing. There exist already
investigations on dynamical aspects of fission. This line of
development needs also further inspection to learn more about the
limits of presently used methods and possibly to improve on them.

Self-consistent models are optimized by least-squares fits to data.
Trend- and correlations-analysis built on the underlying least-squares
fits shows that the relations of fission properties and forces
parameters and NMP are mixed. We see the reduction of uncertainties
when fixing a couple of NMP. But there is no single NMP which has
predominant influence.  Fission properties depend on every detail of
the parametrisation, the bulk properties given by NMP as well as
pairing strength and spin-orbit coupling (determining shell effects).

All this shows that a proper description of fission is extremely
demanding.
More research is needed to achieve a higher reliability for
predictions of fission lifetimes in the context of self-consistent
mean-field models.

\paragraph{\bf Acknowledgements}

This work was supported by the Bundesministerium f\"ur
Bildung und Forschung (BMBF) under contract number 05P12RFFTB; by
Narodowe Centrum Nauki under grant no. 2011/01/B/ST2/05131. M.K. was
co-financed by LEA COPIGAL funds, and the work of LMR was supported in
part by Spanish MINECO grants Nos. FPA2012-34694 and FIS2012-34479 and
by the Consolider-Ingenio 2010 program MULTIDARK CSD2009-00064.

\bibliography{barriers}
\bibliographystyle{elsarticle-num}
\end{document}